\documentclass[aps,ams,amsmath,prl,longbibliography,twocolumn]{revtex4-1}
\usepackage{graphics}
\usepackage{graphicx}
\usepackage{epstopdf,epsfig}
\usepackage{amsmath}
\usepackage{amssymb}
\usepackage{amsfonts}
\usepackage{mathrsfs}
\usepackage{bm}
\usepackage{xcolor}
\usepackage{bbm}
\usepackage{hyperref}
\usepackage[normalem]{ulem}
\usepackage{comment}
\usepackage{soul}
\usepackage{pgfplots}
\pgfplotsset{compat=1.16}
\usetikzlibrary{arrows.meta}

\begin{document}

\title{Nonreciprocal noise of superconducting fluctuations}

\author{Alex Levchenko}
\affiliation{Department of Physics, University of Wisconsin--Madison, Madison, Wisconsin 53706, USA}

\date{November 27, 2026}

\begin{abstract}
Current noise carries information about the stochastic dynamics of collective degrees of freedom that is not contained in linear transport coefficients. Here we formulate the theory of nonequilibrium noise in noncentrosymmetric superconductors in the fluctuation-dominated regime above the critical temperature. The theory is based on the stochastic time-dependent Ginzburg-Landau equation derived microscopically for such systems. Its relaxation rate and Langevin forces contain kinetic Lifshitz invariants. These are the dynamical counterparts of the Lifshitz invariants of the free energy, which determine the spectrum of fluctuating Cooper pairs. We show that the height of the Josephson noise peak of a voltage-biased tunnel junction depends on the helical wave vector of the pairing state, that the paraconductivity current noise of a current-carrying film is nonreciprocal, and that the odd part of the noise power obeys a nonlinear fluctuation relation with the magnetochiral-anisotropy current. We also find that the third cumulant of the fluctuation current is finite in equilibrium. It is odd in the magnetic field and is determined by the same invariants. We derive the equilibrium counting statistics of the fluctuation current. These effects can be studied by noise spectroscopy and nanoscale magnetometry of two-dimensional noncentrosymmetric superconductors.
\end{abstract}

\maketitle

\emph{Introduction.---}Noise measurements probe the stochastic dynamics of electronic systems and give information that is not available from time-averaged transport \cite{BlanterButtiker,KoganBook,Beenakker:2003}. Shot noise measurements revealed the fractional charge $e/3$ of Laughlin quasiparticles \cite{dePicciotto:1997,Saminadayar:1997}, the charge $e/4$ at the $5/2$ state \cite{Dolev:2008}, and the effective charge $2e$ of Andreev transport \cite{Jehl:2000}. Measurements of the third cumulant \cite{Reulet:2003,Bomze:2005} and real-time electron counting \cite{Gustavsson:2006} led to the development of full counting statistics \cite{LevitovLesovik:1993,Nazarov:2007}. In this framework nonlinear fluctuation relations connect the equilibrium cumulants of odd order to the nonlinear response coefficients of the lower cumulants \cite{Forster:2008,SaitoUtsumi:2008,Nagaev:2010}. The odd cumulants are finite only when time-reversal symmetry is broken.

In superconducting mesoscopic structures, noise has mostly been studied in situations where the condensate is fixed and enters as a boundary condition. Andreev reflection doubles the shot noise and changes its voltage dependence in NS and SNS structures \cite{NagaevButtiker:2001,Nagaev:2001,Cuevas:1999,Bezuglyi:1999,Bezuglyi:2001}, and proximity-induced correlations modify the noise generated in the normal regions \cite{Houzet:2003}. The noise caused by the dynamics of the pairing field in the bulk has been studied less. Such dynamics exists above $T_c$. Superconducting fluctuations have the characteristic Ginzburg-Landau (GL) time $\tau_{\mathrm{GL}}=\pi/8(T-T_c)$ and length $\xi_{\mathrm{GL}}=\sqrt{D\tau_{\mathrm{GL}}}$. Their Langevin dynamics in a driven film was analyzed by Kulik \cite{Kulik:1970} and by Scalapino \cite{Scalapino:1970}, following the discovery of fluctuation conductivity \cite{AL:1968,Maki:1968,Thompson:1970} (see the textbook \cite{LarkinVarlamov:2005} for the comprehensive coverage). One prediction of this kind is that the current-noise spectrum of a voltage-biased tunnel junction has a peak at the Josephson frequency $\omega_J=2eV$ in the normal state. This peak was observed in pair-field susceptibility experiments \cite{AndersonGoldman:1970,Bergeal:2008}. Early theoretical studies of this effect addressed narrow contacts, where only temporal fluctuations of the order parameter need to be retained \cite{MartinBalatsky:2000,Dai:2000}, as well as mesoscopic rings \cite{DorinFistul:1992,Shimshoni:1993}. For planar junctions, where the spatial dispersion of the pair field is essential, the noise spectrum was derived systematically from the Keldysh nonlinear sigma model (NLSM) in Ref.~\cite{Levchenko:2008}, with extensions to counting statistics \cite{BagretsLevchenko:2014} and to the relaxation-resolved noise mechanisms of two-dimensional films \cite{Nagaev:1991,Kwak:2023}. Two developments have renewed interest in this physics. First, nitrogen-vacancy (NV) magnetometry now senses current noise noninvasively at the nanoscale \cite{Kolkowitz:2015}, and covariance magnetometry resolves nonlocal, two-point fluctuation correlations \cite{Rovny:2022}, which are the quantities already computed in recent works \cite{Orgad:2026,Cheng:2026,Romare:2026}. Second, many two-dimensional superconductors (e.g. gated MoS$_2$ \cite{Ye:2012,Lu:2015,Saito:2016}, NbSe$_2$ \cite{Xi:2016}, SrTiO$_3$-based systems \cite{Itahashi:2020}) are noncentrosymmetric. The spectrum of their fluctuating pairs is nonreciprocal in the pair momentum. This is known to produce magnetochiral anisotropy (MCA) and supercurrent diode effects (SDE) in nonlinear transport \cite{Rikken:2001,Wakatsuki:2017,Itahashi:2020,TokuraNagaosa:2018,Ando:2020,Nadeem:2023,Shaffer:2025}. This raises the question of how broken inversion symmetry affects the noise of fluctuating Cooper pairs.

We answer this question using the extended stochastic time-dependent Ginzburg-Landau (TDGL) theory derived recently from the NLSM \cite{FLS:2000,Yurkevich:2001,LevchenkoKamenev:2007,KamenevBook,Liao:2017,VBT:2021,VBT:2022,BergeretTokatly:2014} for a disordered Rashba superconductor in an in-plane Zeeman field \cite{Liu:2026}. The result of that work used here is the existence of kinetic Lifshitz invariants. These are terms odd in momentum in the relaxation rate of each fluctuation mode. By the fluctuation-dissipation theorem (FDT) the same terms appear in the Langevin noise. They are the dynamical counterparts of the Lifshitz invariants of the free energy \cite{MineevSamokhin:1994,DimitrovaFeigelman:2007,Kaur:2005,Edelstein:1995,IlicBergeret:2022,Hasan:2024,Hasan:2025}. A nonreciprocal pair spectrum does not by itself lead to nonreciprocal noise. We find three restrictions. The helical shift, which is the largest nonreciprocal term of the spectrum, does not contribute to the field-odd noise of a film. A junction biased by voltage alone has no noise odd in the field. The remaining field-odd noise satisfies the nonlinear fluctuation relations \cite{Forster:2008,SaitoUtsumi:2008,Nagaev:2010}.

\emph{Fluctuation propagators.---}Consider a two-dimensional film with Rashba spin-orbit coupling of strength $\alpha$ (polar axis $\hat{\bm z}$ normal to the plane), in-plane Zeeman field $\bm h$, diffusion constant $D$, and elastic scattering time $\tau$ ($\hbar=k_B=1$; $\epsilon=(T-T_c)/T$). All nonreciprocal terms depend on the in-plane polar vector $\bm n=\bm h\times\bm\alpha$. Both electrodes of the junction considered below are films of this type, and all results refer to the Gaussian fluctuation regime above $T_c$, $\mathrm{Gi}\ll\epsilon\ll1$, where the pair field has no condensate and its dynamics is linear. The retarded pair propagator derived in Ref.~\cite{Liu:2026} reads
\begin{equation}
\mathcal L_R^{-1}(\bm q,\Omega)
=-\frac{\pi}{8T}\Big[\varepsilon(\bm q)-i\Omega\big(1+\bm u\cdot\bm q\big)\Big],
\label{eq:LR}
\end{equation}
where $\bm u=u\,\hat{\bm n}$ is the kinetic Lifshitz invariant and the spectrum of the fluctuation-induced Cooper pairs is
\begin{equation}
\varepsilon(\bm q)=\tau^{-1}_{\mathrm{GL}}
+D\,(\bm q-\bm q_0)^2
+\Xi\,(\hat{\bm n}\cdot\bm q)\,q^2 .
\label{eq:spectrum}
\end{equation}
The spectrum \eqref{eq:spectrum} is nonreciprocal in two ways. The helical shift $\bm q_0=q_0\hat{\bm n}$ moves the minimum of the spectrum to a finite momentum. The cubic Lifshitz term $\Xi$ makes the dispersion asymmetric about this minimum, so that pairs with momenta along $+\hat{\bm n}$ and $-\hat{\bm n}$ remain inequivalent after the shift is removed by a change of variables. The kinetic invariant $u$ makes the relaxation rate depend on the direction of the momentum. Fluctuations with momenta along $+\bm n$ and $-\bm n$ decay at different rates. Onsager reciprocity allows such a term because $\bm h$ is odd under time reversal. For the Rashba model all three parameters were derived microscopically in Ref.~\cite{Liu:2026}; when the Dyakonov-Perel spin-relaxation rate is small compared to temperature they read
\begin{equation}
u=\frac{(\alpha p_F\tau)^2\alpha h}{2T^2},
\quad
\Xi=D\,u,
\quad
q_0=\frac{14\zeta(3)}{\pi^3}\,\frac{(\alpha p_F\tau)^2\alpha h}{T\,D}.
\label{eq:params}
\end{equation}
In this regime the cubic and kinetic invariants are related by $\Xi=Du$. This relation is exact at any spin-relaxation rate, because both coefficients are given by the same derivative of the singlet-triplet kernel. In contrast, $q_0$ crosses over to the disorder-independent value $4\alpha h/v_F^2$ at strong relaxation \cite{Liu:2026}. Field-odd observables require one more step. Equation~(\ref{eq:spectrum}) omits the even term of order $q^4$, and Eq.~(\ref{eq:LR}) omits the even term of order $q^2$ in the friction coefficient. Both are small in $\epsilon$ near $T_c$. However, the shift $\bm q\to\bm q+\bm q_0$, which removes the helical term, generates from them corrections to $\Xi$ and $u$ of the same order as the values in Eq.~(\ref{eq:params}). As a result, field-odd observables contain the reduced couplings $u_{\rm eff}=R\,u$ and $\Xi_{\rm eff}=R\,\Xi$, and the relation $\Xi_{\rm eff}=Du_{\rm eff}$ remains valid \cite{Liu:2026}. In the weak-relaxation regime $R=1-392\zeta(3)^2/\pi^6\simeq0.41$. At strong relaxation $R\simeq(2/\pi^2)(\ln g)/g$ with $g=\Gamma_t/4\pi T$. Below, $u$ and $\Xi$ denote the weak-relaxation values (\ref{eq:params}), and the factor $R$ is written explicitly. At arbitrary $\Gamma_t/T$ the combination $(\alpha p_F\tau)^2\alpha h/T^2$ in Eqs.~(\ref{eq:dSw}), (\ref{eq:Nyquist}), and (\ref{eq:S3}) is replaced by $2u$. In equilibrium the Keldysh component is determined by the FDT, $\mathcal L_K^{-1}=\coth(\tfrac{\Omega}{2T})[\mathcal L_R^{-1}-\mathcal L_A^{-1}]$. The pair correlator that enters the noise calculations is then
\begin{equation}
\mathcal L_K(\bm q,\Omega)\big|_{\Omega\ll T}
=-\frac{32iT^2}{\pi}\,
\frac{1+\bm u\cdot\bm q}{\varepsilon^2(\bm q)+\Omega^2(1+\bm u\cdot\bm q)^2}.
\label{eq:LK}
\end{equation}
It depends on the direction of $\bm q$ for each mode, because the Langevin force acting on a fluctuation with momentum $\bm q$ has intensity proportional to $1+\bm u\cdot\bm q$. The question is whether this asymmetry remains in measurable noise spectra, which involve an integral over all pair momenta. Symmetry gives a simple rule. After the momentum integration a spectrum can be odd in $\bm h$ only if the measurement involves a second in-plane polar vector, which forms a scalar product with $\bm n$.

\begin{figure*}[t]
\centering
\begin{tikzpicture}[scale=0.92,>=Stealth]
% ---------------- panel (a): tunnel junction ----------------
\begin{scope}[shift={(0,0)}]
\node at (-2.1,1.55) {(a)};
\draw[fill=blue!10,draw=blue!50] (-1.9,0.55) rectangle (1.9,1.15);
\draw[fill=gray!30,draw=gray!60] (-1.9,0.35) rectangle (1.9,0.55);
\draw[fill=red!10,draw=red!50] (-1.9,-0.25) rectangle (1.9,0.35);
\node[blue!60!black] at (-1.15,0.87) {\footnotesize SC$_L$};
\node[red!60!black] at (-1.15,0.06) {\footnotesize SC$_R$};
\draw[->,very thick,blue!70!black] (0.15,0.85) -- (1.35,0.85) node[midway,above] {\footnotesize $\bm q_0^L$};
\draw[->,very thick,red!70!black] (1.35,0.05) -- (0.15,0.05) node[midway,below] {\footnotesize $\bm q_0^R$};
\draw[thick] (2.15,0.85) -- (2.6,0.85) -- (2.6,0.05) -- (2.15,0.05);
\node at (2.95,0.45) {\footnotesize $V$};
\draw (2.5,0.32) circle (0.14);
\node at (0,-0.85) {\footnotesize $S_J$ peaks at $\omega_J=2eV$;\; height $\propto P(\kappa)$};
\end{scope}
% ---------------- panel (b): film + NV ----------------
\begin{scope}[shift={(7.6,0)}]
\node at (-2.1,1.55) {(b)};
\draw[fill=red!8,draw=red!40] (-1.9,-0.35) -- (1.5,-0.35) -- (2.3,0.45) -- (-1.1,0.45) -- cycle;
\draw[->,very thick] (-0.9,0.05) -- (0.9,0.05) node[midway,below] {\footnotesize $\bm j\parallel\bm n$};
\draw[->,thick] (1.15,-0.15) -- (1.95,-0.15) node[right] {\footnotesize $\bm h$};
\node at (-1.45,0.22) {$\bm\alpha\,\odot$};
% NV sensor
\draw[fill=violet!20,draw=violet!70,rotate around={45:(0.4,1.25)}] (0.25,1.1) rectangle (0.55,1.4);
\node[violet!70!black] at (1.15,1.32) {\footnotesize NV};
\draw[dashed] (0.4,0.5) -- (0.4,1.02);
\node at (0.72,0.72) {\footnotesize $d$};
\node at (0.2,-0.95) {\footnotesize $\delta S(\omega)\propto \bm E\cdot\bm n$;\;\;$S_3\neq0$ at $E=0$};
\end{scope}
% ---------------- panel (c): F curves ----------------
\begin{scope}[shift={(0.2,-5.6)}]
\node at (-2.3,2.35) {(c)};
\begin{axis}[width=6.4cm,height=4.6cm,
xlabel={$\Omega=\omega\tau_{\mathrm{GL}}$},ylabel={$F(\Omega)$},
xmin=0,xmax=16,ymin=0,ymax=1.02,
label style={font=\footnotesize},tick label style={font=\footnotesize},
legend style={font=\scriptsize},grid=both,grid style={gray!15}]
\addplot[blue,very thick] coordinates {(0.0,1.00000) (0.25,0.98970) (0.5,0.96019) (0.75,0.91514) (1.0,0.85941) (1.5,0.73458) (2.0,0.61371) (2.5,0.50915) (3.0,0.42325) (4.0,0.29882) (5.0,0.21857) (6.0,0.16537) (8.0,0.10287) (10.0,0.06957) (13.0,0.04312) (16.0,0.02921)};
\addlegendentry{$F_u$ (kinetic)}
\addplot[red,dashed,very thick] coordinates {(0.0,1.00000) (0.25,0.99483) (0.5,0.97979) (0.75,0.95617) (1.0,0.92574) (1.5,0.85222) (2.0,0.77259) (2.5,0.69531) (3.0,0.62445) (4.0,0.50590) (5.0,0.41535) (6.0,0.34632) (8.0,0.25129) (10.0,0.19107) (13.0,0.13520) (16.0,0.10124)};
\addlegendentry{$F_\Xi$ (cubic)}
\addplot[black,densely dotted,thick] coordinates {(0.0,1.00000) (0.25,0.99227) (0.5,0.96999) (0.75,0.93565) (1.0,0.89257) (1.5,0.79340) (2.0,0.69315) (2.5,0.60223) (3.0,0.52385) (4.0,0.40236) (5.0,0.31696) (6.0,0.25584) (8.0,0.17708) (10.0,0.13032) (13.0,0.08916) (16.0,0.06522)};
\addlegendentry{$2N(\Omega)$ (sum$/2$)}
\end{axis}
\end{scope}
% ---------------- panel (d): P(kappa) ----------------
\begin{scope}[shift={(7.8,-5.6)}]
\node at (-2.3,2.35) {(d)};
\begin{axis}[width=6.4cm,height=4.6cm,
xlabel={$\kappa=\delta q_0\,\xi_{\mathrm{GL}}$},ylabel={$P(\kappa)$},
xmin=0,xmax=3,ymin=0,ymax=1.02,
label style={font=\footnotesize},tick label style={font=\footnotesize},
legend style={font=\scriptsize},grid=both,grid style={gray!15}]
\addplot[blue,very thick] coordinates {(0.0,1.0000) (0.1,0.9835) (0.2,0.9367) (0.3,0.8660) (0.4,0.7804) (0.5,0.6886) (0.6,0.5977) (0.8,0.4360) (1.0,0.3116) (1.2,0.2221) (1.5,0.1359) (1.8,0.0859) (2.2,0.0493) (2.6,0.0300) (3.0,0.0192)};
\addlegendentry{$P(\kappa)$, Eq.~(\ref{eq:P})}
\addplot[red,dashed,thick,domain=0:0.55] {1-(5/3)*x^2};
\addlegendentry{$1-\tfrac53\kappa^2$}
\end{axis}
\end{scope}
\end{tikzpicture}
\caption{(a) Voltage-biased tunnel junction of two noncentrosymmetric superconducting films above $T_c$. An in-plane field induces helical shifts $\bm q_0^{L,R}$ of the fluctuation spectra; for antiparallel polar axes the shifts are opposite, and the Josephson noise peak at $\omega_J=2eV$ measures the mismatch through $P(\kappa)$. (b) Current-biased noncentrosymmetric film with a nanoscale NV magnetometer: the fluctuation current noise acquires a component odd in both $\bm j$ and $\bm h$, and the equilibrium third cumulant is finite. (c) Universal spectral shapes of the two contributions to the nonreciprocal noise, Eq.~(\ref{eq:dSw}); the contributions of the kinetic and cubic Lifshitz invariants have different frequency dependence on the scale $1/\tau_{\mathrm{GL}}$, while their sum, Eq.~(\ref{eq:Fsum}), is the Josephson line shape $N(\Omega)$ (dotted). (d) Helix-mismatch suppression of the Josephson noise peak, Eq.~(\ref{eq:P}), with the small-$\kappa$ expansion.}
\label{fig:1}
\end{figure*}
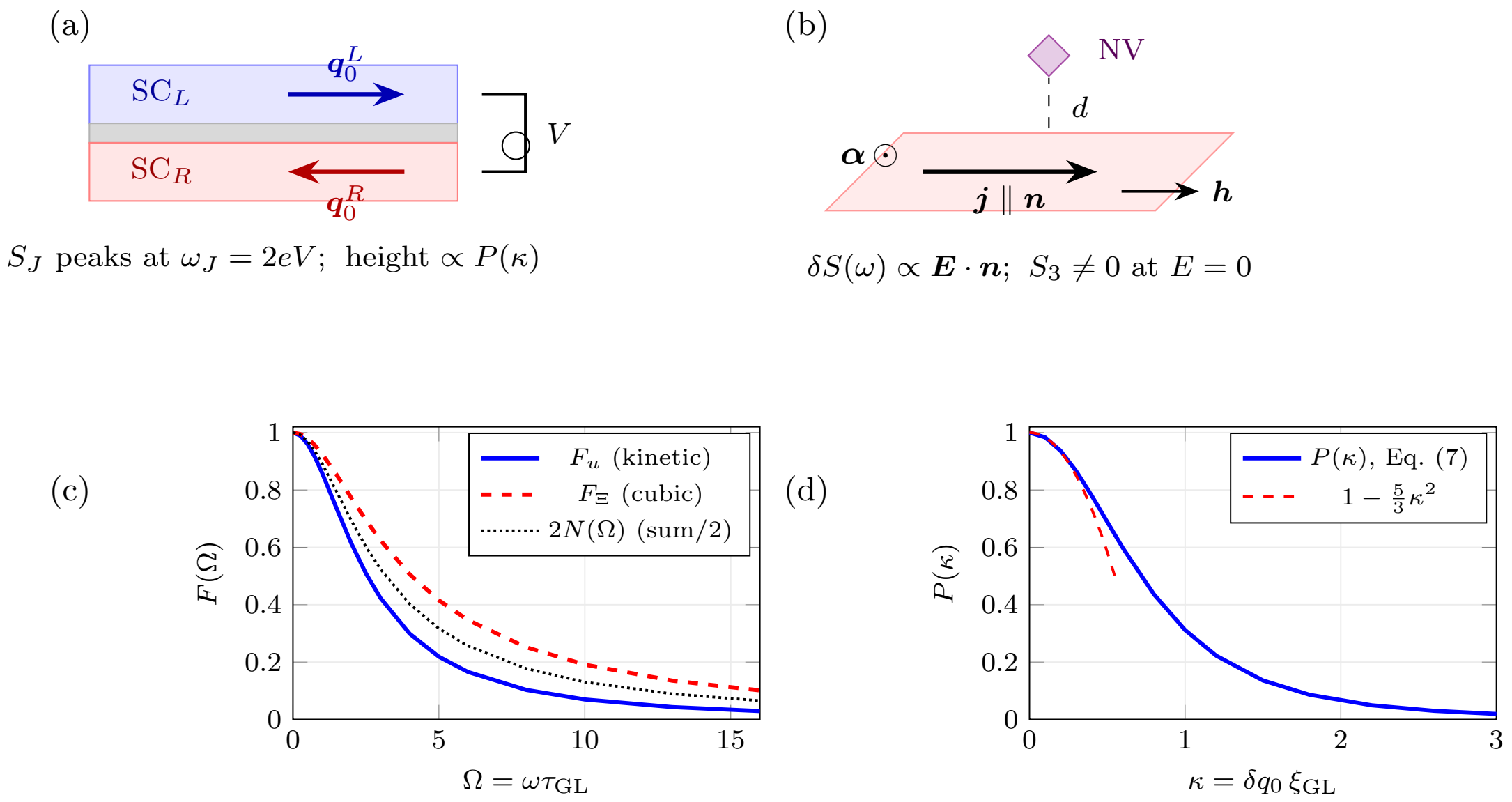

\emph{Josephson noise and the helical wave vector.---}Consider first the voltage-biased planar tunnel junction of two such films [Fig.~\ref{fig:1}(a)], the geometry whose noise spectrum above $T_c$ was derived from the Keldysh NLSM in Ref.~\cite{Levchenko:2008}. The fluctuational Josephson current $I_J\propto\!\int\! d^2r\,[\Delta_R\Delta_L^*e^{-i\omega_Jt}-\mathrm{c.c.}]$ has zero average and a finite variance. Averaging the product of two currents with the correlator (\ref{eq:LK}) for each electrode gives
\begin{equation}
S_J(\omega)=\sum_\pm\frac{C_J}{8\pi^2}
\!\int\! d^2q\,
\frac{\varepsilon_L+\varepsilon_R}
{\varepsilon_L\varepsilon_R\big[(\varepsilon_L+\varepsilon_R)^2+\omega_\pm^2\big]},
\label{eq:SJ}
\end{equation}
with $\omega_\pm=\omega\pm\omega_J$, $C_J=16\,g_T^2T_c^2/(e^2\nu_L\nu_R\mathcal A)$, $g_T$ the tunnel conductance, $\mathcal A$ the junction area, and $\varepsilon_{L,R}=\varepsilon_{L,R}(\bm q)$ the electrode dispersions of Eq.~(\ref{eq:LR}). In a planar junction that conserves the in-plane momentum both propagators are taken at the same $\bm q$. The factors $1+\bm u_a\cdot\bm q$ contribute only at second order and are omitted. For identical centrosymmetric electrodes, Eq.~(\ref{eq:SJ}) reproduces the line shape of Ref.~\cite{Levchenko:2008},
\begin{equation}
S_J(\omega)=\sum_\pm\frac{\pi \mathrm{Gi}^2}{64\,T_c}
\Big(\frac{g_TT_c}{e}\Big)^{\!2}\frac{\xi_0^2}{\mathcal A}
\Big(\frac{T_c}{T-T_c}\Big)^{\!2}N(\omega_\pm\tau_{\mathrm{GL}};\kappa).
\label{eq:SJfinal}
\end{equation}
Noncentrosymmetric electrodes add one parameter, $\kappa=\delta q_0\,\xi_{\mathrm{GL}}$, where $\delta\bm q_0=(\bm q_0^L-\bm q_0^R)/2$ is half the difference of the helical wave vectors of the two electrodes. Here $\mathrm{Gi}=1/\nu D$, $\xi_0^2=D/T_c$, and $N(z;0)=(4/z^2)\ln\sqrt{1+z^2/4}$. We first note a symmetry constraint. Every nonreciprocal coupling enters through $\bm n\cdot\bm q$, so each propagator obeys $\mathcal L(\bm q;\bm h)=\mathcal L(-\bm q;-\bm h)$. The change of variable $\bm q\to-\bm q$ in Eq.~(\ref{eq:SJ}) then gives $S(\omega;\bm h)=S(\omega;-\bm h)$. This relation is exact and holds for any orientation of the polar axes of the two films. A uniform junction biased by voltage alone therefore has no noise odd in $\bm h$, because the voltage is a scalar and does not provide a second polar vector. The effect that remains is even in $\bm h$ and changes when the polar axis of one film is reversed. The common part of $\bm q_0^{L}$ and $\bm q_0^{R}$ is removed by a shift of the integration variable, and only their difference enters. At the peak,
\begin{equation}
\frac{N(0;\kappa)}{N(0;0)}\equiv P(\kappa)=\frac{2}{\pi}\!\int\!\!\frac{d^2k\;\big[1+k^2+\kappa^2\big]^{-1}}
{\big[1+(\bm k-\bm\kappa)^2\big]\big[1+(\bm k+\bm\kappa)^2\big]},
\label{eq:P}
\end{equation}
with $P=1-\tfrac53\kappa^2+O(\kappa^4)$ and $P\simeq\ln(\kappa^2)/2\kappa^4$ at $\kappa\gg1$ [Fig.~\ref{fig:1}(d)]. For identical films with parallel polar axes $\delta q_0=0$ and the peak is not affected. For antiparallel axes $\delta q_0=q_0$ and $\kappa=q_0\,\xi_{\mathrm{GL}}$ grows as $\epsilon^{-1/2}$ on approach to $T_c$ at fixed field. Within the Gaussian regime, $\epsilon\gtrsim\mathrm{Gi}$, $\kappa$ is bounded by $q_0\xi_0/\sqrt{\mathrm{Gi}}$. For fields such that $q_0(h)\xi_0\gtrsim\sqrt{\mathrm{Gi}}$ the peak of the antiparallel junction is strongly suppressed at $\epsilon\lesssim(q_0\xi_0)^2$, while that of the parallel junction is unchanged. For weaker fields the suppression is perturbative, $1-P\simeq\tfrac53 q_0^2\xi_{\mathrm{GL}}^2$. An in-plane field $\bm H$ also shifts the pair momenta of the two electrodes relative to each other by $q_{\rm orb}=2\pi Hd_{\rm eff}/\Phi_0$, where $d_{\rm eff}$ is their center-to-center distance. This is the orbital (Fraunhofer) effect, which is present in centrosymmetric junctions as well. The orbital shift is collinear with $\delta\bm q_0$, does not depend on disorder, does not change when a polar axis is reversed, and is typically larger than $q_0$ (SM). Equation~(\ref{eq:P}) then applies with $\kappa=|q_{\rm orb}/2\pm q_0|\,\xi_{\mathrm{GL}}$, where the two signs correspond to the two antiparallel arrangements of the polar axes. The difference of the two peak heights is $P_+-P_-\simeq-\tfrac{10}{3}\xi_{\mathrm{GL}}^2q_{\rm orb}q_0$. It is linear in $q_0$ because of the cross term with the orbital shift. A measurement of this difference gives $q_0(h,T,\tau)$, including its dependence on disorder \cite{Liu:2026,IlicBergeret:2022,VBT:2022}, in the regime above $T_c$ where no supercurrent flows.

\emph{Nonreciprocal noise of a driven film.---}According to the rule above, noise odd in $\bm h$ requires an in-plane vector. The simplest choice is a transport current [Fig.~\ref{fig:1}(b)]. The fluctuation (paraconductivity) current of a film is $j=\int_{\bm q}J(\bm q)|\Delta_{\bm q}|^2$. The current vertex is $J=2e\nu\,\partial_{q_n}a$, where $a(\bm q)=(\pi/8T)\,\varepsilon(\bm q)$ is the Ginzburg-Landau kernel. Each mode obeys Langevin dynamics with the relaxation rate $\lambda_{\bm q}=\varepsilon(\bm q)/(1+\bm u\cdot\bm q)$ and the noise intensity of Eq.~(\ref{eq:LK}). The FDT relates the two, so that the equilibrium occupation is $\bar n_{\bm q}=T/\nu a(\bm q)$. In a dc field $\bm E\parallel\hat{\bm n}$ each mode drifts, $\dot{\bm q}=2e\bm E$, and the stationary occupation follows from the kinetic equation $2e\bm E\cdot\partial_{\bm q}n_{\bm q}=-2\lambda_{\bm q}[n_{\bm q}-\bar n_{\bm q}]$. The relaxation rate is taken at the instantaneous momentum of the mode \cite{Liu:2026}. We compute the current-noise spectrum to linear order in $E$. The calculation takes into account the change of the current vertex and of the relaxation rate as the momentum of a mode drifts [Supplemental Material (SM)]. The result is
\begin{equation}
\delta S_{\mathrm{odd}}(\omega)=R\,\frac{3\pi e^3E}{256}\,
\frac{(\alpha p_F\tau)^2\alpha h}{T^2\epsilon^{2}}
\Big[F_u(\Omega)+F_\Xi(\Omega)\Big],
\label{eq:dSw}
\end{equation}
with $\Omega=\omega\tau_{\mathrm{GL}}$. This spectrum is odd in both $E$ and $\bm h$. The noise power of the film is therefore different for the two directions of the current. Its frequency dependence is set by $\tau_{\mathrm{GL}}$ and consists of two universal functions [Fig.~\ref{fig:1}(c)]. The function $F_u$ is the contribution of the kinetic invariant $\bm u$, that is, of the nonreciprocal Langevin forces,
\begin{equation}
F_u(\Omega)=\frac{8}{\Omega^4}\Big[\Omega^2-4\ln\!\Big(1+\frac{\Omega^2}{4}\Big)\Big].
\label{eq:Fu}
\end{equation}
It decays as $8/\Omega^2$ at large $\Omega$. The function $F_\Xi$ is the contribution of the cubic Lifshitz invariant,
\begin{equation}
F_\Xi(\Omega)=\frac{8}{\Omega^4}
\Big[\big(\Omega^2+4\big)\ln\!\Big(1+\frac{\Omega^2}{4}\Big)-\Omega^2\Big].
\label{eq:FXi}
\end{equation}
It decays more slowly, as $(8/\Omega^2)\ln(\Omega^2/4)$. At $\omega=0$ the two contributions are equal; this holds in any dimension. The terms linear in $\bm q_0$ cancel at all frequencies, because the helical shift can be removed by a change of the integration variable. The two contributions have different frequency dependence. In the Rashba model $\Xi=Du$, and their sum is
\begin{equation}
F_u(\Omega)+F_\Xi(\Omega)=\frac{8}{\Omega^2}\ln\Big(1+\frac{\Omega^2}{4}\Big)=4N(\Omega).
\label{eq:Fsum}
\end{equation}
The field-odd spectrum is then proportional to $N(\omega\tau_{\mathrm{GL}})$, the function that describes the Josephson noise peak in Eq.~(\ref{eq:SJfinal}), and contains no additional parameters. The relation $\Xi=Du$ follows from the microscopic model and not from symmetry. In a general noncentrosymmetric TDGL theory $u$ and $\Xi$ are independent, and the line shape deviates from $N(\Omega)$ by an amount proportional to $\Xi/Du-1$. A measurement of the frequency dependence therefore tests this relation, which is not possible with dc measurements. A contribution to the noise linear in $E$ is also possible in a centrosymmetric superconductor, but it requires particle-hole asymmetry, as discussed in Ref.~\cite{Orgad:2026}. The contribution found here has a different origin. It does not require particle-hole asymmetry and is caused by broken inversion and time-reversal symmetries, which enter through the Lifshitz invariants.

At zero frequency Eq.~(\ref{eq:dSw}) is related to nonlinear transport. The same theory gives the rectified current of the fluctuation MCA \cite{Liu:2026}. The two quantities satisfy
\begin{equation}
\delta S_{\mathrm{odd}}(0)=3T\,\frac{\partial\,\delta j_{\mathrm{MCA}}}{\partial E},
\;\;
\delta j_{\mathrm{MCA}}=R\frac{\pi e^3E^2}{256T}
\frac{(\alpha p_F\tau)^2\alpha h}{T^2\epsilon^{2}}.
\label{eq:Nyquist}
\end{equation}
Equation~(\ref{eq:Nyquist}) is the nonlinear fluctuation relation $\partial_ES=3T\,\partial_E^2j$ between the field-odd noise susceptibility and the field-odd nonlinear conductance \cite{SaitoUtsumi:2008,Nagaev:2010}. In the present theory it follows from the calculation. The factor $3T$ has the following origin. The change of the stationary occupation contributes $2T\partial_E\delta j_{\mathrm{MCA}}$ to the noise, as one integration by parts shows. The drift of the current vertex and of the relaxation rate adds one half of this amount (SM). Equation~(\ref{eq:Nyquist}) therefore holds in any dimension, for any mode dispersion, and at any order in the nonreciprocal couplings. It is the analog of the Johnson-Nyquist relation at the first nonlinear order: to linear order in the field each mode remains in local equilibrium at its instantaneous momentum. Equation~(\ref{eq:Nyquist}) refers to the paraconductivity (Aslamazov-Larkin) current. The Maki-Thompson and inelastic-relaxation contributions \cite{Kwak:2023} are discussed in the SM.

\emph{Third cumulant without bias, and counting statistics.---}The second cumulant of the current is even under time reversal, so its field-odd part requires a drive. The third cumulant is odd under time reversal and can be finite in equilibrium in a magnetic field. For Gaussian fluctuations the modes are independent, and the current is a sum of squares of Ornstein-Uhlenbeck processes. The zero-frequency third cumulant of the fluctuation current is given by momentum integrals of the same type as in Eq.~(\ref{eq:dSw}) (SM):
\begin{equation}
S_3^{\mathrm{eq}}
=3\!\int\!\!\frac{d^2q}{(2\pi)^2}\frac{J^3\bar n^3}{\lambda^2}
=R\,\frac{3\pi e^3T}{64}\,
\frac{(\alpha p_F\tau)^2\,\alpha h}{T^2\,\epsilon^{2}} .
\label{eq:S3}
\end{equation}
The distribution of the equilibrium current fluctuations of a noncentrosymmetric superconducting film above $T_c$ is therefore asymmetric, and the asymmetry changes sign under $\bm h\to-\bm h$. A finite equilibrium $S_3$ requires broken time-reversal and inversion symmetries. This case is described by the nonlinear fluctuation relations \cite{Forster:2008,SaitoUtsumi:2008,Nagaev:2010}. With $S_1=\partial_E\delta S_{\mathrm{odd}}(0)$ and $G_2=\partial_E^2\delta j_{\mathrm{MCA}}$, the equilibrium third cumulant satisfies $S_3^{\mathrm{eq}}=3TS_1-3T^2G_2$ \cite{Forster:2008}. If the generating function has the symmetry given below Eq.~(\ref{eq:CGF}), the stronger relations $S_3^{\mathrm{eq}}=2TS_1=6T^2G_2$ hold \cite{SaitoUtsumi:2008,Nagaev:2010}. Equations~(\ref{eq:Nyquist}) and (\ref{eq:S3}) satisfy both. The same relations were found for a macroscopic Langevin system in Ref.~\cite{Nagaev:2010}. The equilibrium counting statistics can be found in closed form \cite{CameronMartin:1945,JakemanPike:1968}. Let $W=\int_0^{\mathcal T}dt\,j(t)$ be the charge transported by the fluctuation current during a long observation time $\mathcal T$, and let $\eta$ be the counting variable conjugate to it. The corresponding cumulant generating function is a sum over independent modes,
\begin{equation}
\lim_{\mathcal T\to\infty}\frac{\ln\langle e^{i\eta W}\rangle}{\mathcal T}
=\sum_{\bm q}\lambda_{\bm q}
\Big[1-\sqrt{1-\tfrac{2i\eta J(\bm q)\bar n_{\bm q}}{\lambda_{\bm q}}}\Big].
\label{eq:CGF}
\end{equation}
Here the average $\langle\ldots\rangle$ is taken over the realizations of the Langevin forces, and the cumulants are $c_k=(2k-3)!!\sum_{\bm q}J^k\bar n^k\lambda^{1-k}$. The vertex obeys $J(-\bm q;\bm h)=-J(\bm q;-\bm h)$, while $\lambda$ and $\bar n$ do not change under the simultaneous reversal of $\bm q$ and $\bm h$. Therefore all odd cumulants are odd in $\bm h$ and all even cumulants are even in $\bm h$. This property can be tested by real-time counting or by third-moment measurements \cite{Reulet:2003,Bomze:2005,Gustavsson:2006}. By the same change of variable, all cumulants of a tunnel junction biased by voltage alone are even in $\bm h$. This agrees with the zero-dimensional limit \cite{BagretsLevchenko:2014}, where gradient terms are absent. Equation~(\ref{eq:CGF}) describes the charge carried by the fluctuating pairs (the Aslamazov-Larkin current). In the Gaussian regime it is statistically independent of the normal-electron current, whose fluctuations are Gaussian and even in $\bm h$. The generating function in the presence of the field $E$ is given in the SM. It reproduces Eqs.~(\ref{eq:Nyquist}) and (\ref{eq:S3}) and obeys the symmetry $\mathcal F(\chi;E;\bm h)=\mathcal F(-\chi-E/T;E;-\bm h)$ with $\chi=i\eta$, from which the fluctuation relations above follow \cite{SaitoUtsumi:2008}.

\emph{Outlook.---}The symmetry constraint also indicates how field-odd noise can be obtained in a junction. An in-plane current through one electrode provides the required vector. It changes the height of the Josephson noise peak at $\omega=2eV$ by a relative amount of order $(eE_\parallel/T\epsilon)(\alpha p_F\tau)^2\alpha h/T^2$, and the sign of this change follows the direction of the current. The results above are obtained for the paraconductivity contribution in the Gaussian regime, $\mathrm{Gi}\ll\epsilon\ll1$, at classical frequencies, $\{\omega,\omega_J\}\ll T$, and at weak drive, $2eE\tau_{\mathrm{GL}}\xi_{\mathrm{GL}}\ll1$. Slow quasiparticle relaxation is expected to increase the nonreciprocal noise through the Maki-Thompson \cite{JTM:2026} and correlated Boltzmann-Langevin contributions \cite{Kwak:2023}, with the parameters $\tau_\phi/\tau_{\mathrm{GL}}$ and $T_c\tau_R$, respectively. The predicted noise can be measured by NV magnetometry \cite{Kolkowitz:2015}. The field-odd noise of the film, Eqs.~(\ref{eq:dSw})--(\ref{eq:S3}), has the characteristic frequency $\tau_{\mathrm{GL}}^{-1}=8(T-T_c)/\pi$, which can be brought into the frequency range of NV sensors by approaching $T_c$. A sensor at height $d$ above the film measures current-density correlations at in-plane wave vectors $Q\lesssim1/d$. For $d\gtrsim\xi_{\mathrm{GL}}$ the results for the uniform current apply with corrections of order $\xi_{\mathrm{GL}}^2/d^2$. The sign reversal of the signal with $\bm E$ and with $\bm h$ holds at any $Q$. Covariance magnetometry \cite{Rovny:2022} measures two-point correlations at finite $Q$, where the momentum asymmetry of Eq.~(\ref{eq:LK}) is not removed by the integration. In summary, fluctuating Cooper pairs in noncentrosymmetric superconductors show three nonreciprocal noise effects: a noise power that depends on the direction of the current, a Josephson noise peak that depends on the relative orientation of the polar axes, and counting statistics with cumulants odd in the magnetic field.

\emph{Acknowledgments.---}
I am grateful to Ilya Esterlis for carefully reading the manuscript and providing valuable comments. This work was supported by the U.S. Department of Energy (DOE), Office of Science, Basic Energy Sciences (BES) under Award No. DE-SC0020313. The author acknowledges H. I. Romnes Faculty Fellowship provided by the University of Wisconsin-Madison Office of the Vice Chancellor for Research and Graduate Education with funding from the Wisconsin Alumni Research Foundation. 

\emph{AI use disclosure.---} 
This work in part employed the large language model Claude (Anthropic, model Claude Fable 5.2) for manuscript preparation in particular for symbolic verification
of analytical results. The TDGL analysis follows the author's earlier work on the topic. The author conceived and directed the project, verified the results, and takes full responsibility for the content of the paper.

\bibliography{biblio-NoiseSC}

%merlin.mbs apsrev4-1.bst 2010-07-25 4.21a (PWD, AO, DPC) hacked
%Control: key (0)
%Control: author (0) dotless jnrlst
%Control: editor formatted (1) identically to author
%Control: production of article title (0) allowed
%Control: page (1) range
%Control: year (0) verbatim
%Control: production of eprint (0) enabled
\begin{thebibliography}{72}%
\makeatletter
\providecommand \@ifxundefined [1]{%
 \@ifx{#1\undefined}
}%
\providecommand \@ifnum [1]{%
 \ifnum #1\expandafter \@firstoftwo
 \else \expandafter \@secondoftwo
 \fi
}%
\providecommand \@ifx [1]{%
 \ifx #1\expandafter \@firstoftwo
 \else \expandafter \@secondoftwo
 \fi
}%
\providecommand \natexlab [1]{#1}%
\providecommand \enquote  [1]{``#1''}%
\providecommand \bibnamefont  [1]{#1}%
\providecommand \bibfnamefont [1]{#1}%
\providecommand \citenamefont [1]{#1}%
\providecommand \href@noop [0]{\@secondoftwo}%
\providecommand \href [0]{\begingroup \@sanitize@url \@href}%
\providecommand \@href[1]{\@@startlink{#1}\@@href}%
\providecommand \@@href[1]{\endgroup#1\@@endlink}%
\providecommand \@sanitize@url [0]{\catcode `\\12\catcode `\$12\catcode
  `\&12\catcode `\#12\catcode `\^12\catcode `\_12\catcode `\%12\relax}%
\providecommand \@@startlink[1]{}%
\providecommand \@@endlink[0]{}%
\providecommand \url  [0]{\begingroup\@sanitize@url \@url }%
\providecommand \@url [1]{\endgroup\@href {#1}{\urlprefix }}%
\providecommand \urlprefix  [0]{URL }%
\providecommand \Eprint [0]{\href }%
\providecommand \doibase [0]{http://dx.doi.org/}%
\providecommand \selectlanguage [0]{\@gobble}%
\providecommand \bibinfo  [0]{\@secondoftwo}%
\providecommand \bibfield  [0]{\@secondoftwo}%
\providecommand \translation [1]{[#1]}%
\providecommand \BibitemOpen [0]{}%
\providecommand \bibitemStop [0]{}%
\providecommand \bibitemNoStop [0]{.\EOS\space}%
\providecommand \EOS [0]{\spacefactor3000\relax}%
\providecommand \BibitemShut  [1]{\csname bibitem#1\endcsname}%
\let\auto@bib@innerbib\@empty
%</preamble>
\bibitem [{\citenamefont {Blanter}\ and\ \citenamefont
  {B{\"u}ttiker}(2000)}]{BlanterButtiker}%
  \BibitemOpen
  \bibfield  {author} {\bibinfo {author} {\bibfnamefont {Ya.~M.}\ \bibnamefont
  {Blanter}}\ and\ \bibinfo {author} {\bibfnamefont {M.}~\bibnamefont
  {B{\"u}ttiker}},\ }\bibfield  {title} {\enquote {\bibinfo {title} {Shot noise
  in mesoscopic conductors},}\ }\href@noop {} {\bibfield  {journal} {\bibinfo
  {journal} {Phys. Rep.}\ }\textbf {\bibinfo {volume} {336}},\ \bibinfo {pages}
  {1} (\bibinfo {year} {2000})}\BibitemShut {NoStop}%
\bibitem [{\citenamefont {Kogan}(1996)}]{KoganBook}%
  \BibitemOpen
  \bibfield  {author} {\bibinfo {author} {\bibfnamefont {Sh.}\ \bibnamefont
  {Kogan}},\ }\href@noop {} {\emph {\bibinfo {title} {Electronic Noise and
  Fluctuations in Solids}}}\ (\bibinfo  {publisher} {Cambridge University
  Press},\ \bibinfo {address} {Cambridge},\ \bibinfo {year} {1996})\BibitemShut
  {NoStop}%
\bibitem [{\citenamefont {Beenakker}\ and\ \citenamefont
  {Sch{\"o}nenberger}(2003)}]{Beenakker:2003}%
  \BibitemOpen
  \bibfield  {author} {\bibinfo {author} {\bibfnamefont {C.}~\bibnamefont
  {Beenakker}}\ and\ \bibinfo {author} {\bibfnamefont {C.}~\bibnamefont
  {Sch{\"o}nenberger}},\ }\bibfield  {title} {\enquote {\bibinfo {title}
  {Quantum shot noise},}\ }\href@noop {} {\bibfield  {journal} {\bibinfo
  {journal} {Phys. Today}\ }\textbf {\bibinfo {volume} {56}},\ \bibinfo {pages}
  {37} (\bibinfo {year} {2003})}\BibitemShut {NoStop}%
\bibitem [{\citenamefont {de~Picciotto}\ \emph {et~al.}(1997)\citenamefont
  {de~Picciotto}, \citenamefont {Reznikov}, \citenamefont {Heiblum},
  \citenamefont {Umansky}, \citenamefont {Bunin},\ and\ \citenamefont
  {Mahalu}}]{dePicciotto:1997}%
  \BibitemOpen
  \bibfield  {author} {\bibinfo {author} {\bibfnamefont {R.}~\bibnamefont
  {de~Picciotto}}, \bibinfo {author} {\bibfnamefont {M.}~\bibnamefont
  {Reznikov}}, \bibinfo {author} {\bibfnamefont {M.}~\bibnamefont {Heiblum}},
  \bibinfo {author} {\bibfnamefont {V.}~\bibnamefont {Umansky}}, \bibinfo
  {author} {\bibfnamefont {G.}~\bibnamefont {Bunin}}, \ and\ \bibinfo {author}
  {\bibfnamefont {D.}~\bibnamefont {Mahalu}},\ }\bibfield  {title} {\enquote
  {\bibinfo {title} {Direct observation of a fractional charge},}\ }\href@noop
  {} {\bibfield  {journal} {\bibinfo  {journal} {Nature}\ }\textbf {\bibinfo
  {volume} {389}},\ \bibinfo {pages} {162} (\bibinfo {year}
  {1997})}\BibitemShut {NoStop}%
\bibitem [{\citenamefont {Saminadayar}\ \emph {et~al.}(1997)\citenamefont
  {Saminadayar}, \citenamefont {Glattli}, \citenamefont {Jin},\ and\
  \citenamefont {Etienne}}]{Saminadayar:1997}%
  \BibitemOpen
  \bibfield  {author} {\bibinfo {author} {\bibfnamefont {L.}~\bibnamefont
  {Saminadayar}}, \bibinfo {author} {\bibfnamefont {D.~C.}\ \bibnamefont
  {Glattli}}, \bibinfo {author} {\bibfnamefont {Y.}~\bibnamefont {Jin}}, \ and\
  \bibinfo {author} {\bibfnamefont {B.}~\bibnamefont {Etienne}},\ }\bibfield
  {title} {\enquote {\bibinfo {title} {Observation of the $e/3$ fractionally
  charged {Laughlin} quasiparticle},}\ }\href@noop {} {\bibfield  {journal}
  {\bibinfo  {journal} {Phys. Rev. Lett.}\ }\textbf {\bibinfo {volume} {79}},\
  \bibinfo {pages} {2526} (\bibinfo {year} {1997})}\BibitemShut {NoStop}%
\bibitem [{\citenamefont {Dolev}\ \emph {et~al.}(2008)\citenamefont {Dolev},
  \citenamefont {Heiblum}, \citenamefont {Umansky}, \citenamefont {Stern},\
  and\ \citenamefont {Mahalu}}]{Dolev:2008}%
  \BibitemOpen
  \bibfield  {author} {\bibinfo {author} {\bibfnamefont {M.}~\bibnamefont
  {Dolev}}, \bibinfo {author} {\bibfnamefont {M.}~\bibnamefont {Heiblum}},
  \bibinfo {author} {\bibfnamefont {V.}~\bibnamefont {Umansky}}, \bibinfo
  {author} {\bibfnamefont {Ady}\ \bibnamefont {Stern}}, \ and\ \bibinfo
  {author} {\bibfnamefont {D.}~\bibnamefont {Mahalu}},\ }\bibfield  {title}
  {\enquote {\bibinfo {title} {Observation of a quarter of an electron charge
  at the $\nu=5/2$ quantum {Hall} state},}\ }\href@noop {} {\bibfield
  {journal} {\bibinfo  {journal} {Nature}\ }\textbf {\bibinfo {volume} {452}},\
  \bibinfo {pages} {829} (\bibinfo {year} {2008})}\BibitemShut {NoStop}%
\bibitem [{\citenamefont {Jehl}\ \emph {et~al.}(2000)\citenamefont {Jehl},
  \citenamefont {Sanquer}, \citenamefont {Calemczuk},\ and\ \citenamefont
  {Mailly}}]{Jehl:2000}%
  \BibitemOpen
  \bibfield  {author} {\bibinfo {author} {\bibfnamefont {X.}~\bibnamefont
  {Jehl}}, \bibinfo {author} {\bibfnamefont {M.}~\bibnamefont {Sanquer}},
  \bibinfo {author} {\bibfnamefont {R.}~\bibnamefont {Calemczuk}}, \ and\
  \bibinfo {author} {\bibfnamefont {D.}~\bibnamefont {Mailly}},\ }\bibfield
  {title} {\enquote {\bibinfo {title} {Detection of doubled shot noise in short
  normal-metal/superconductor junctions},}\ }\href@noop {} {\bibfield
  {journal} {\bibinfo  {journal} {Nature}\ }\textbf {\bibinfo {volume} {405}},\
  \bibinfo {pages} {50} (\bibinfo {year} {2000})}\BibitemShut {NoStop}%
\bibitem [{\citenamefont {Reulet}\ \emph {et~al.}(2003)\citenamefont {Reulet},
  \citenamefont {Senzier},\ and\ \citenamefont {Prober}}]{Reulet:2003}%
  \BibitemOpen
  \bibfield  {author} {\bibinfo {author} {\bibfnamefont {B.}~\bibnamefont
  {Reulet}}, \bibinfo {author} {\bibfnamefont {J.}~\bibnamefont {Senzier}}, \
  and\ \bibinfo {author} {\bibfnamefont {D.~E.}\ \bibnamefont {Prober}},\
  }\bibfield  {title} {\enquote {\bibinfo {title} {Environmental effects in the
  third moment of voltage fluctuations in a tunnel junction},}\ }\href@noop {}
  {\bibfield  {journal} {\bibinfo  {journal} {Phys. Rev. Lett.}\ }\textbf
  {\bibinfo {volume} {91}},\ \bibinfo {pages} {196601} (\bibinfo {year}
  {2003})}\BibitemShut {NoStop}%
\bibitem [{\citenamefont {Bomze}\ \emph {et~al.}(2005)\citenamefont {Bomze},
  \citenamefont {Gershon}, \citenamefont {Shovkun}, \citenamefont {Levitov},\
  and\ \citenamefont {Reznikov}}]{Bomze:2005}%
  \BibitemOpen
  \bibfield  {author} {\bibinfo {author} {\bibfnamefont {Yu.}\ \bibnamefont
  {Bomze}}, \bibinfo {author} {\bibfnamefont {G.}~\bibnamefont {Gershon}},
  \bibinfo {author} {\bibfnamefont {D.}~\bibnamefont {Shovkun}}, \bibinfo
  {author} {\bibfnamefont {L.~S.}\ \bibnamefont {Levitov}}, \ and\ \bibinfo
  {author} {\bibfnamefont {M.}~\bibnamefont {Reznikov}},\ }\bibfield  {title}
  {\enquote {\bibinfo {title} {Measurement of counting statistics of electron
  transport in a tunnel junction},}\ }\href@noop {} {\bibfield  {journal}
  {\bibinfo  {journal} {Phys. Rev. Lett.}\ }\textbf {\bibinfo {volume} {95}},\
  \bibinfo {pages} {176601} (\bibinfo {year} {2005})}\BibitemShut {NoStop}%
\bibitem [{\citenamefont {Gustavsson}\ \emph {et~al.}(2006)\citenamefont
  {Gustavsson}, \citenamefont {Leturcq}, \citenamefont {Simovi{\v c}},
  \citenamefont {Schleser}, \citenamefont {Ihn}, \citenamefont {Studerus},
  \citenamefont {Ensslin}, \citenamefont {Driscoll},\ and\ \citenamefont
  {Gossard}}]{Gustavsson:2006}%
  \BibitemOpen
  \bibfield  {author} {\bibinfo {author} {\bibfnamefont {S.}~\bibnamefont
  {Gustavsson}}, \bibinfo {author} {\bibfnamefont {R.}~\bibnamefont {Leturcq}},
  \bibinfo {author} {\bibfnamefont {B.}~\bibnamefont {Simovi{\v c}}}, \bibinfo
  {author} {\bibfnamefont {R.}~\bibnamefont {Schleser}}, \bibinfo {author}
  {\bibfnamefont {T.}~\bibnamefont {Ihn}}, \bibinfo {author} {\bibfnamefont
  {P.}~\bibnamefont {Studerus}}, \bibinfo {author} {\bibfnamefont
  {K.}~\bibnamefont {Ensslin}}, \bibinfo {author} {\bibfnamefont {D.~C.}\
  \bibnamefont {Driscoll}}, \ and\ \bibinfo {author} {\bibfnamefont {A.~C.}\
  \bibnamefont {Gossard}},\ }\bibfield  {title} {\enquote {\bibinfo {title}
  {Counting statistics of single electron transport in a quantum dot},}\
  }\href@noop {} {\bibfield  {journal} {\bibinfo  {journal} {Phys. Rev. Lett.}\
  }\textbf {\bibinfo {volume} {96}},\ \bibinfo {pages} {076605} (\bibinfo
  {year} {2006})}\BibitemShut {NoStop}%
\bibitem [{\citenamefont {Levitov}\ and\ \citenamefont
  {Lesovik}(1993)}]{LevitovLesovik:1993}%
  \BibitemOpen
  \bibfield  {author} {\bibinfo {author} {\bibfnamefont {L.~S.}\ \bibnamefont
  {Levitov}}\ and\ \bibinfo {author} {\bibfnamefont {G.~B.}\ \bibnamefont
  {Lesovik}},\ }\bibfield  {title} {\enquote {\bibinfo {title} {Charge
  distribution in quantum shot noise},}\ }\href@noop {} {\bibfield  {journal}
  {\bibinfo  {journal} {Pis'ma Zh. Eksp. Teor. Fiz.}\ }\textbf {\bibinfo
  {volume} {58}},\ \bibinfo {pages} {225} (\bibinfo {year} {1993})},\ \bibinfo
  {note} {[JETP Lett. {\bf 58}, 230 (1993)]}\BibitemShut {NoStop}%
\bibitem [{\citenamefont {Nazarov}(2007)}]{Nazarov:2007}%
  \BibitemOpen
  \bibfield  {author} {\bibinfo {author} {\bibfnamefont {Y.V.}\ \bibnamefont
  {Nazarov}},\ }\bibfield  {title} {\enquote {\bibinfo {title} {Full counting
  statistics and field theory},}\ }\href {\doibase
  https://doi.org/10.1002/andp.200751910-1106} {\bibfield  {journal} {\bibinfo
  {journal} {Annalen der Physik}\ }\textbf {\bibinfo {volume} {519}},\ \bibinfo
  {pages} {720--735} (\bibinfo {year} {2007})}\BibitemShut {NoStop}%
\bibitem [{\citenamefont {F{\"o}rster}\ and\ \citenamefont
  {B{\"u}ttiker}(2008)}]{Forster:2008}%
  \BibitemOpen
  \bibfield  {author} {\bibinfo {author} {\bibfnamefont {H.}~\bibnamefont
  {F{\"o}rster}}\ and\ \bibinfo {author} {\bibfnamefont {M.}~\bibnamefont
  {B{\"u}ttiker}},\ }\bibfield  {title} {\enquote {\bibinfo {title}
  {Fluctuation relations without microreversibility in nonlinear transport},}\
  }\href {\doibase 10.1103/PhysRevLett.101.136805} {\bibfield  {journal}
  {\bibinfo  {journal} {Phys. Rev. Lett.}\ }\textbf {\bibinfo {volume} {101}},\
  \bibinfo {pages} {136805} (\bibinfo {year} {2008})}\BibitemShut {NoStop}%
\bibitem [{\citenamefont {Saito}\ and\ \citenamefont
  {Utsumi}(2008)}]{SaitoUtsumi:2008}%
  \BibitemOpen
  \bibfield  {author} {\bibinfo {author} {\bibfnamefont {K.}~\bibnamefont
  {Saito}}\ and\ \bibinfo {author} {\bibfnamefont {Y.}~\bibnamefont {Utsumi}},\
  }\bibfield  {title} {\enquote {\bibinfo {title} {Symmetry in full counting
  statistics, fluctuation theorem, and relations among nonlinear transport
  coefficients in the presence of a magnetic field},}\ }\href {\doibase
  10.1103/PhysRevB.78.115429} {\bibfield  {journal} {\bibinfo  {journal} {Phys.
  Rev. B}\ }\textbf {\bibinfo {volume} {78}},\ \bibinfo {pages} {115429}
  (\bibinfo {year} {2008})}\BibitemShut {NoStop}%
\bibitem [{\citenamefont {Nagaev}\ \emph {et~al.}(2010)\citenamefont {Nagaev},
  \citenamefont {Ayvazyan}, \citenamefont {Sergeeva},\ and\ \citenamefont
  {B{\"u}ttiker}}]{Nagaev:2010}%
  \BibitemOpen
  \bibfield  {author} {\bibinfo {author} {\bibfnamefont {K.~E.}\ \bibnamefont
  {Nagaev}}, \bibinfo {author} {\bibfnamefont {O.~S.}\ \bibnamefont
  {Ayvazyan}}, \bibinfo {author} {\bibfnamefont {N.~Yu.}\ \bibnamefont
  {Sergeeva}}, \ and\ \bibinfo {author} {\bibfnamefont {M.}~\bibnamefont
  {B{\"u}ttiker}},\ }\bibfield  {title} {\enquote {\bibinfo {title}
  {Magnetic-field-induced non-{Gaussian} fluctuations in macroscopic
  equilibrium systems},}\ }\href {\doibase 10.1103/PhysRevLett.105.146802}
  {\bibfield  {journal} {\bibinfo  {journal} {Phys. Rev. Lett.}\ }\textbf
  {\bibinfo {volume} {105}},\ \bibinfo {pages} {146802} (\bibinfo {year}
  {2010})}\BibitemShut {NoStop}%
\bibitem [{\citenamefont {Nagaev}\ and\ \citenamefont
  {B{\"u}ttiker}(2001)}]{NagaevButtiker:2001}%
  \BibitemOpen
  \bibfield  {author} {\bibinfo {author} {\bibfnamefont {K.~E.}\ \bibnamefont
  {Nagaev}}\ and\ \bibinfo {author} {\bibfnamefont {M.}~\bibnamefont
  {B{\"u}ttiker}},\ }\bibfield  {title} {\enquote {\bibinfo {title}
  {Semiclassical theory of shot noise in disordered
  superconductor--normal-metal contacts},}\ }\href@noop {} {\bibfield
  {journal} {\bibinfo  {journal} {Phys. Rev. B}\ }\textbf {\bibinfo {volume}
  {63}},\ \bibinfo {pages} {081301(R)} (\bibinfo {year} {2001})}\BibitemShut
  {NoStop}%
\bibitem [{\citenamefont {Nagaev}(2001)}]{Nagaev:2001}%
  \BibitemOpen
  \bibfield  {author} {\bibinfo {author} {\bibfnamefont {K.~E.}\ \bibnamefont
  {Nagaev}},\ }\bibfield  {title} {\enquote {\bibinfo {title}
  {Frequency-dependent shot noise in long disordered
  superconductor--normal-metal--superconductor contacts},}\ }\href@noop {}
  {\bibfield  {journal} {\bibinfo  {journal} {Phys. Rev. Lett.}\ }\textbf
  {\bibinfo {volume} {86}},\ \bibinfo {pages} {3112} (\bibinfo {year}
  {2001})}\BibitemShut {NoStop}%
\bibitem [{\citenamefont {Cuevas}\ \emph {et~al.}(1999)\citenamefont {Cuevas},
  \citenamefont {Mart{\'\i}n-Rodero},\ and\ \citenamefont
  {Levy~Yeyati}}]{Cuevas:1999}%
  \BibitemOpen
  \bibfield  {author} {\bibinfo {author} {\bibfnamefont {J.~C.}\ \bibnamefont
  {Cuevas}}, \bibinfo {author} {\bibfnamefont {A.}~\bibnamefont
  {Mart{\'\i}n-Rodero}}, \ and\ \bibinfo {author} {\bibfnamefont
  {A.}~\bibnamefont {Levy~Yeyati}},\ }\bibfield  {title} {\enquote {\bibinfo
  {title} {Shot noise and coherent multiple charge transfer in superconducting
  quantum point contacts},}\ }\href@noop {} {\bibfield  {journal} {\bibinfo
  {journal} {Phys. Rev. Lett.}\ }\textbf {\bibinfo {volume} {82}},\ \bibinfo
  {pages} {4086} (\bibinfo {year} {1999})}\BibitemShut {NoStop}%
\bibitem [{\citenamefont {Bezuglyi}\ \emph {et~al.}(1999)\citenamefont
  {Bezuglyi}, \citenamefont {Bratus'}, \citenamefont {Shumeiko},\ and\
  \citenamefont {Wendin}}]{Bezuglyi:1999}%
  \BibitemOpen
  \bibfield  {author} {\bibinfo {author} {\bibfnamefont {E.~V.}\ \bibnamefont
  {Bezuglyi}}, \bibinfo {author} {\bibfnamefont {E.~N.}\ \bibnamefont
  {Bratus'}}, \bibinfo {author} {\bibfnamefont {V.~S.}\ \bibnamefont
  {Shumeiko}}, \ and\ \bibinfo {author} {\bibfnamefont {G.}~\bibnamefont
  {Wendin}},\ }\bibfield  {title} {\enquote {\bibinfo {title} {Multiple
  {Andreev} reflections and enhanced shot noise in diffusive
  superconducting-normal-superconductor junctions},}\ }\href {\doibase
  10.1103/PhysRevLett.83.2050} {\bibfield  {journal} {\bibinfo  {journal}
  {Phys. Rev. Lett.}\ }\textbf {\bibinfo {volume} {83}},\ \bibinfo {pages}
  {2050--2053} (\bibinfo {year} {1999})}\BibitemShut {NoStop}%
\bibitem [{\citenamefont {Bezuglyi}\ \emph {et~al.}(2001)\citenamefont
  {Bezuglyi}, \citenamefont {Bratus'}, \citenamefont {Shumeiko},\ and\
  \citenamefont {Wendin}}]{Bezuglyi:2001}%
  \BibitemOpen
  \bibfield  {author} {\bibinfo {author} {\bibfnamefont {E.~V.}\ \bibnamefont
  {Bezuglyi}}, \bibinfo {author} {\bibfnamefont {E.~N.}\ \bibnamefont
  {Bratus'}}, \bibinfo {author} {\bibfnamefont {V.~S.}\ \bibnamefont
  {Shumeiko}}, \ and\ \bibinfo {author} {\bibfnamefont {G.}~\bibnamefont
  {Wendin}},\ }\bibfield  {title} {\enquote {\bibinfo {title} {Current noise in
  long diffusive {SNS} junctions in the incoherent multiple {Andreev}
  reflections regime},}\ }\href {\doibase 10.1103/PhysRevB.63.100501}
  {\bibfield  {journal} {\bibinfo  {journal} {Phys. Rev. B}\ }\textbf {\bibinfo
  {volume} {63}},\ \bibinfo {pages} {100501(R)} (\bibinfo {year}
  {2001})}\BibitemShut {NoStop}%
\bibitem [{\citenamefont {Houzet}\ and\ \citenamefont
  {Mineev}(2003)}]{Houzet:2003}%
  \BibitemOpen
  \bibfield  {author} {\bibinfo {author} {\bibfnamefont {M.}~\bibnamefont
  {Houzet}}\ and\ \bibinfo {author} {\bibfnamefont {V.~P.}\ \bibnamefont
  {Mineev}},\ }\bibfield  {title} {\enquote {\bibinfo {title} {Influence of the
  proximity effect on the shot noise in a diffusive metallic wire},}\ }\href
  {\doibase 10.1103/PhysRevB.67.184524} {\bibfield  {journal} {\bibinfo
  {journal} {Phys. Rev. B}\ }\textbf {\bibinfo {volume} {67}},\ \bibinfo
  {pages} {184524} (\bibinfo {year} {2003})}\BibitemShut {NoStop}%
\bibitem [{\citenamefont {Kulik}(1970)}]{Kulik:1970}%
  \BibitemOpen
  \bibfield  {author} {\bibinfo {author} {\bibfnamefont {I.~O.}\ \bibnamefont
  {Kulik}},\ }\bibfield  {title} {\enquote {\bibinfo {title} {Nonstationary
  effects in the resistive state of superconducting films},}\ }\href@noop {}
  {\bibfield  {journal} {\bibinfo  {journal} {Zh. Eksp. Teor. Fiz.}\ }\textbf
  {\bibinfo {volume} {59}},\ \bibinfo {pages} {584} (\bibinfo {year} {1970})},\
  \bibinfo {note} {[Sov. Phys. JETP {\bf 32}, 318 (1971)]}\BibitemShut
  {NoStop}%
\bibitem [{\citenamefont {Scalapino}(1970)}]{Scalapino:1970}%
  \BibitemOpen
  \bibfield  {author} {\bibinfo {author} {\bibfnamefont {D.~J.}\ \bibnamefont
  {Scalapino}},\ }\bibfield  {title} {\enquote {\bibinfo {title} {Pair
  tunneling as a probe of fluctuations in superconductors},}\ }\href@noop {}
  {\bibfield  {journal} {\bibinfo  {journal} {Phys. Rev. Lett.}\ }\textbf
  {\bibinfo {volume} {24}},\ \bibinfo {pages} {1052} (\bibinfo {year}
  {1970})}\BibitemShut {NoStop}%
\bibitem [{\citenamefont {Aslamazov}\ and\ \citenamefont
  {Larkin}(1968)}]{AL:1968}%
  \BibitemOpen
  \bibfield  {author} {\bibinfo {author} {\bibfnamefont {L.~G.}\ \bibnamefont
  {Aslamazov}}\ and\ \bibinfo {author} {\bibfnamefont {A.~I.}\ \bibnamefont
  {Larkin}},\ }\bibfield  {title} {\enquote {\bibinfo {title} {Effect of
  fluctuations on the properties of a superconductor above the critical
  temperature},}\ }\href@noop {} {\bibfield  {journal} {\bibinfo  {journal}
  {Fiz. Tverd. Tela}\ }\textbf {\bibinfo {volume} {10}},\ \bibinfo {pages}
  {1104} (\bibinfo {year} {1968})},\ \bibinfo {note} {[Sov. Phys. Solid State
  {\bf 10}, 875 (1968)]}\BibitemShut {NoStop}%
\bibitem [{\citenamefont {Maki}(1968)}]{Maki:1968}%
  \BibitemOpen
  \bibfield  {author} {\bibinfo {author} {\bibfnamefont {Kazumi}\ \bibnamefont
  {Maki}},\ }\bibfield  {title} {\enquote {\bibinfo {title} {The critical
  fluctuation of the order parameter in type-{II} superconductors},}\ }\href
  {\doibase 10.1143/PTP.39.897} {\bibfield  {journal} {\bibinfo  {journal}
  {Progress of Theoretical Physics}\ }\textbf {\bibinfo {volume} {39}},\
  \bibinfo {pages} {897--906} (\bibinfo {year} {1968})}\BibitemShut {NoStop}%
\bibitem [{\citenamefont {Thompson}(1970)}]{Thompson:1970}%
  \BibitemOpen
  \bibfield  {author} {\bibinfo {author} {\bibfnamefont {Richard~S.}\
  \bibnamefont {Thompson}},\ }\bibfield  {title} {\enquote {\bibinfo {title}
  {Microwave, flux flow, and fluctuation resistance of dirty type-{II}
  superconductors},}\ }\href {\doibase 10.1103/PhysRevB.1.327} {\bibfield
  {journal} {\bibinfo  {journal} {Phys. Rev. B}\ }\textbf {\bibinfo {volume}
  {1}},\ \bibinfo {pages} {327--333} (\bibinfo {year} {1970})}\BibitemShut
  {NoStop}%
\bibitem [{\citenamefont {Larkin}\ and\ \citenamefont
  {Varlamov}(2005)}]{LarkinVarlamov:2005}%
  \BibitemOpen
  \bibfield  {author} {\bibinfo {author} {\bibfnamefont {A.~I.}\ \bibnamefont
  {Larkin}}\ and\ \bibinfo {author} {\bibfnamefont {A.~A.}\ \bibnamefont
  {Varlamov}},\ }\href@noop {} {\emph {\bibinfo {title} {Theory of Fluctuations
  in Superconductors}}}\ (\bibinfo  {publisher} {Clarendon Press},\ \bibinfo
  {address} {Oxford},\ \bibinfo {year} {2005})\BibitemShut {NoStop}%
\bibitem [{\citenamefont {Anderson}\ and\ \citenamefont
  {Goldman}(1971)}]{AndersonGoldman:1970}%
  \BibitemOpen
  \bibfield  {author} {\bibinfo {author} {\bibfnamefont {J.~T.}\ \bibnamefont
  {Anderson}}\ and\ \bibinfo {author} {\bibfnamefont {A.~M.}\ \bibnamefont
  {Goldman}},\ }\bibfield  {title} {\enquote {\bibinfo {title} {Experimental
  determination of the pair susceptibility of a superconductor},}\ }\href
  {\doibase 10.1103/PhysRevLett.26.534} {\bibfield  {journal} {\bibinfo
  {journal} {Phys. Rev. Lett.}\ }\textbf {\bibinfo {volume} {26}},\ \bibinfo
  {pages} {534} (\bibinfo {year} {1971})}\BibitemShut {NoStop}%
\bibitem [{\citenamefont {Bergeal}\ \emph {et~al.}(2008)\citenamefont
  {Bergeal}, \citenamefont {Lesueur}, \citenamefont {Aprili}, \citenamefont
  {Faini}, \citenamefont {Contour},\ and\ \citenamefont
  {Leridon}}]{Bergeal:2008}%
  \BibitemOpen
  \bibfield  {author} {\bibinfo {author} {\bibfnamefont {N.}~\bibnamefont
  {Bergeal}}, \bibinfo {author} {\bibfnamefont {J.}~\bibnamefont {Lesueur}},
  \bibinfo {author} {\bibfnamefont {M.}~\bibnamefont {Aprili}}, \bibinfo
  {author} {\bibfnamefont {G.}~\bibnamefont {Faini}}, \bibinfo {author}
  {\bibfnamefont {J.~P.}\ \bibnamefont {Contour}}, \ and\ \bibinfo {author}
  {\bibfnamefont {B.}~\bibnamefont {Leridon}},\ }\bibfield  {title} {\enquote
  {\bibinfo {title} {Pairing fluctuations in the pseudogap state of
  copper-oxide superconductors probed by the {Josephson} effect},}\ }\href@noop
  {} {\bibfield  {journal} {\bibinfo  {journal} {Nat. Phys.}\ }\textbf
  {\bibinfo {volume} {4}},\ \bibinfo {pages} {608} (\bibinfo {year}
  {2008})}\BibitemShut {NoStop}%
\bibitem [{\citenamefont {Martin}\ and\ \citenamefont
  {Balatsky}(2000)}]{MartinBalatsky:2000}%
  \BibitemOpen
  \bibfield  {author} {\bibinfo {author} {\bibfnamefont {Ivar}\ \bibnamefont
  {Martin}}\ and\ \bibinfo {author} {\bibfnamefont {Alexander}\ \bibnamefont
  {Balatsky}},\ }\bibfield  {title} {\enquote {\bibinfo {title} {Probing
  pseudogap by {Josephson} tunneling},}\ }\href {\doibase
  10.1103/PhysRevB.62.R6124} {\bibfield  {journal} {\bibinfo  {journal} {Phys.
  Rev. B}\ }\textbf {\bibinfo {volume} {62}},\ \bibinfo {pages}
  {R6124(R)--R6126(R)} (\bibinfo {year} {2000})}\BibitemShut {NoStop}%
\bibitem [{\citenamefont {Dai}\ \emph {et~al.}(2000)\citenamefont {Dai},
  \citenamefont {Xiang}, \citenamefont {Ng},\ and\ \citenamefont
  {Su}}]{Dai:2000}%
  \BibitemOpen
  \bibfield  {author} {\bibinfo {author} {\bibfnamefont {Xi}~\bibnamefont
  {Dai}}, \bibinfo {author} {\bibfnamefont {Tao}\ \bibnamefont {Xiang}},
  \bibinfo {author} {\bibfnamefont {Tai-Kai}\ \bibnamefont {Ng}}, \ and\
  \bibinfo {author} {\bibfnamefont {Zhao-bin}\ \bibnamefont {Su}},\ }\bibfield
  {title} {\enquote {\bibinfo {title} {Probing superconducting phase
  fluctuations from the current noise spectrum of pseudogapped
  metal-superconductor tunnel junctions},}\ }\href {\doibase
  10.1103/PhysRevLett.85.3009} {\bibfield  {journal} {\bibinfo  {journal}
  {Phys. Rev. Lett.}\ }\textbf {\bibinfo {volume} {85}},\ \bibinfo {pages}
  {3009--3012} (\bibinfo {year} {2000})}\BibitemShut {NoStop}%
\bibitem [{\citenamefont {Dorin}\ and\ \citenamefont
  {Fistul'}(1992)}]{DorinFistul:1992}%
  \BibitemOpen
  \bibfield  {author} {\bibinfo {author} {\bibfnamefont {V.~V.}\ \bibnamefont
  {Dorin}}\ and\ \bibinfo {author} {\bibfnamefont {M.~V.}\ \bibnamefont
  {Fistul'}},\ }\bibfield  {title} {\enquote {\bibinfo {title} {Critical
  current of a {Josephson} junction with spatial fluctuations of the order
  parameter},}\ }\href {\doibase 10.1103/PhysRevB.46.13951} {\bibfield
  {journal} {\bibinfo  {journal} {Phys. Rev. B}\ }\textbf {\bibinfo {volume}
  {46}},\ \bibinfo {pages} {13951--13955} (\bibinfo {year} {1992})}\BibitemShut
  {NoStop}%
\bibitem [{\citenamefont {Shimshoni}\ \emph {et~al.}(1993)\citenamefont
  {Shimshoni}, \citenamefont {Goldbart},\ and\ \citenamefont
  {Goldenfeld}}]{Shimshoni:1993}%
  \BibitemOpen
  \bibfield  {author} {\bibinfo {author} {\bibfnamefont {Efrat}\ \bibnamefont
  {Shimshoni}}, \bibinfo {author} {\bibfnamefont {Paul~M.}\ \bibnamefont
  {Goldbart}}, \ and\ \bibinfo {author} {\bibfnamefont {Nigel}\ \bibnamefont
  {Goldenfeld}},\ }\bibfield  {title} {\enquote {\bibinfo {title} {Josephson
  interference phenomena above {${\mathit{T}}_{\mathit{c}}$}},}\ }\href
  {\doibase 10.1103/PhysRevB.48.9865} {\bibfield  {journal} {\bibinfo
  {journal} {Phys. Rev. B}\ }\textbf {\bibinfo {volume} {48}},\ \bibinfo
  {pages} {9865--9868} (\bibinfo {year} {1993})}\BibitemShut {NoStop}%
\bibitem [{\citenamefont {Levchenko}(2008)}]{Levchenko:2008}%
  \BibitemOpen
  \bibfield  {author} {\bibinfo {author} {\bibfnamefont {A.}~\bibnamefont
  {Levchenko}},\ }\bibfield  {title} {\enquote {\bibinfo {title} {Josephson
  current noise above {$T_c$} in superconducting tunnel junctions},}\
  }\href@noop {} {\bibfield  {journal} {\bibinfo  {journal} {Phys. Rev. B}\
  }\textbf {\bibinfo {volume} {78}},\ \bibinfo {pages} {104507} (\bibinfo
  {year} {2008})}\BibitemShut {NoStop}%
\bibitem [{\citenamefont {Bagrets}\ and\ \citenamefont
  {Levchenko}(2014)}]{BagretsLevchenko:2014}%
  \BibitemOpen
  \bibfield  {author} {\bibinfo {author} {\bibfnamefont {D.}~\bibnamefont
  {Bagrets}}\ and\ \bibinfo {author} {\bibfnamefont {A.}~\bibnamefont
  {Levchenko}},\ }\bibfield  {title} {\enquote {\bibinfo {title}
  {Nonequilibrium noise and current fluctuations at the superconducting phase
  transition},}\ }\href@noop {} {\bibfield  {journal} {\bibinfo  {journal}
  {Phys. Rev. B}\ }\textbf {\bibinfo {volume} {90}},\ \bibinfo {pages}
  {180505(R)} (\bibinfo {year} {2014})}\BibitemShut {NoStop}%
\bibitem [{\citenamefont {Nagaev}(1991)}]{Nagaev:1991}%
  \BibitemOpen
  \bibfield  {author} {\bibinfo {author} {\bibfnamefont {K.E.}\ \bibnamefont
  {Nagaev}},\ }\bibfield  {title} {\enquote {\bibinfo {title} {Theory of excess
  noise in superconductors above {$T_c$}},}\ }\href {\doibase
  https://doi.org/10.1016/0921-4534(91)91512-3} {\bibfield  {journal} {\bibinfo
   {journal} {Physica C: Superconductivity and its Applications}\ }\textbf
  {\bibinfo {volume} {184}},\ \bibinfo {pages} {149--158} (\bibinfo {year}
  {1991})}\BibitemShut {NoStop}%
\bibitem [{\citenamefont {Kwak}\ \emph {et~al.}(2023)\citenamefont {Kwak},
  \citenamefont {Pellett}, \citenamefont {K{\"o}nig},\ and\ \citenamefont
  {Levchenko}}]{Kwak:2023}%
  \BibitemOpen
  \bibfield  {author} {\bibinfo {author} {\bibfnamefont {J.}~\bibnamefont
  {Kwak}}, \bibinfo {author} {\bibfnamefont {E.}~\bibnamefont {Pellett}},
  \bibinfo {author} {\bibfnamefont {E.~J.}\ \bibnamefont {K{\"o}nig}}, \ and\
  \bibinfo {author} {\bibfnamefont {A.}~\bibnamefont {Levchenko}},\ }\bibfield
  {title} {\enquote {\bibinfo {title} {Fluctuation-driven excess noise near
  superconducting phase transition},}\ }\href@noop {} {\bibfield  {journal}
  {\bibinfo  {journal} {Ann. Phys.}\ }\textbf {\bibinfo {volume} {456}},\
  \bibinfo {pages} {169307} (\bibinfo {year} {2023})}\BibitemShut {NoStop}%
\bibitem [{\citenamefont {Kolkowitz}\ \emph {et~al.}(2015)\citenamefont
  {Kolkowitz}, \citenamefont {Safira}, \citenamefont {High}, \citenamefont
  {Devlin}, \citenamefont {Choi}, \citenamefont {Unterreithmeier},
  \citenamefont {Patterson}, \citenamefont {Zibrov}, \citenamefont
  {Manucharyan}, \citenamefont {Park},\ and\ \citenamefont
  {Lukin}}]{Kolkowitz:2015}%
  \BibitemOpen
  \bibfield  {author} {\bibinfo {author} {\bibfnamefont {S.}~\bibnamefont
  {Kolkowitz}}, \bibinfo {author} {\bibfnamefont {A.}~\bibnamefont {Safira}},
  \bibinfo {author} {\bibfnamefont {A.~A.}\ \bibnamefont {High}}, \bibinfo
  {author} {\bibfnamefont {R.~C.}\ \bibnamefont {Devlin}}, \bibinfo {author}
  {\bibfnamefont {S.}~\bibnamefont {Choi}}, \bibinfo {author} {\bibfnamefont
  {Q.~P.}\ \bibnamefont {Unterreithmeier}}, \bibinfo {author} {\bibfnamefont
  {D.}~\bibnamefont {Patterson}}, \bibinfo {author} {\bibfnamefont {A.~S.}\
  \bibnamefont {Zibrov}}, \bibinfo {author} {\bibfnamefont {V.~E.}\
  \bibnamefont {Manucharyan}}, \bibinfo {author} {\bibfnamefont
  {H.}~\bibnamefont {Park}}, \ and\ \bibinfo {author} {\bibfnamefont {M.~D.}\
  \bibnamefont {Lukin}},\ }\bibfield  {title} {\enquote {\bibinfo {title}
  {Probing {Johnson} noise and ballistic transport in normal metals with a
  single-spin qubit},}\ }\href@noop {} {\bibfield  {journal} {\bibinfo
  {journal} {Science}\ }\textbf {\bibinfo {volume} {347}},\ \bibinfo {pages}
  {1129} (\bibinfo {year} {2015})}\BibitemShut {NoStop}%
\bibitem [{\citenamefont {Rovny}\ \emph {et~al.}(2022)\citenamefont {Rovny},
  \citenamefont {Yuan}, \citenamefont {Fitzpatrick}, \citenamefont {Abdalla},
  \citenamefont {Futamura}, \citenamefont {Fox}, \citenamefont {Cambria},
  \citenamefont {Kolkowitz},\ and\ \citenamefont {de~Leon}}]{Rovny:2022}%
  \BibitemOpen
  \bibfield  {author} {\bibinfo {author} {\bibfnamefont {J.}~\bibnamefont
  {Rovny}}, \bibinfo {author} {\bibfnamefont {Z.}~\bibnamefont {Yuan}},
  \bibinfo {author} {\bibfnamefont {M.}~\bibnamefont {Fitzpatrick}}, \bibinfo
  {author} {\bibfnamefont {A.~I.}\ \bibnamefont {Abdalla}}, \bibinfo {author}
  {\bibfnamefont {L.}~\bibnamefont {Futamura}}, \bibinfo {author}
  {\bibfnamefont {C.}~\bibnamefont {Fox}}, \bibinfo {author} {\bibfnamefont
  {M.~C.}\ \bibnamefont {Cambria}}, \bibinfo {author} {\bibfnamefont
  {S.}~\bibnamefont {Kolkowitz}}, \ and\ \bibinfo {author} {\bibfnamefont
  {N.~P.}\ \bibnamefont {de~Leon}},\ }\bibfield  {title} {\enquote {\bibinfo
  {title} {Nanoscale covariance magnetometry with diamond quantum sensors},}\
  }\href@noop {} {\bibfield  {journal} {\bibinfo  {journal} {Science}\ }\textbf
  {\bibinfo {volume} {378}},\ \bibinfo {pages} {1301} (\bibinfo {year}
  {2022})}\BibitemShut {NoStop}%
\bibitem [{\citenamefont {Orgad}(2026)}]{Orgad:2026}%
  \BibitemOpen
  \bibfield  {author} {\bibinfo {author} {\bibfnamefont {Dror}\ \bibnamefont
  {Orgad}},\ }\bibfield  {title} {\enquote {\bibinfo {title} {Signatures of
  {Gaussian} superconducting fluctuations in nonlocal noise magnetometry},}\
  }\href {\doibase 10.1103/x5ym-1w9j} {\bibfield  {journal} {\bibinfo
  {journal} {Phys. Rev. B}\ }\textbf {\bibinfo {volume} {114}},\ \bibinfo
  {pages} {264511} (\bibinfo {year} {2026})}\BibitemShut {NoStop}%
\bibitem [{\citenamefont {Cheng}\ \emph {et~al.}(2026)\citenamefont {Cheng},
  \citenamefont {Kim}, \citenamefont {Diessel}, \citenamefont {Zu},\ and\
  \citenamefont {Chatterjee}}]{Cheng:2026}%
  \BibitemOpen
  \bibfield  {author} {\bibinfo {author} {\bibfnamefont {Jiajie}\ \bibnamefont
  {Cheng}}, \bibinfo {author} {\bibfnamefont {Jaewon}\ \bibnamefont {Kim}},
  \bibinfo {author} {\bibfnamefont {Oriana~K.}\ \bibnamefont {Diessel}},
  \bibinfo {author} {\bibfnamefont {Chong}\ \bibnamefont {Zu}}, \ and\ \bibinfo
  {author} {\bibfnamefont {Shubhayu}\ \bibnamefont {Chatterjee}},\ }\href
  {https://arxiv.org/abs/2602.20265} {\enquote {\bibinfo {title} {Qubit noise
  spectroscopy of superconducting dynamics in a magnetic field},}\ } (\bibinfo
  {year} {2026}),\ \Eprint {http://arxiv.org/abs/2602.20265} {arXiv:2602.20265
  [cond-mat.supr-con]} \BibitemShut {NoStop}%
\bibitem [{\citenamefont {Romare}\ \emph {et~al.}(2026)\citenamefont {Romare},
  \citenamefont {Esterlis}, \citenamefont {Kolkowitz},\ and\ \citenamefont
  {Levchenko}}]{Romare:2026}%
  \BibitemOpen
  \bibfield  {author} {\bibinfo {author} {\bibfnamefont {Gustav}\ \bibnamefont
  {Romare}}, \bibinfo {author} {\bibfnamefont {Ilya}\ \bibnamefont {Esterlis}},
  \bibinfo {author} {\bibfnamefont {Shimon}\ \bibnamefont {Kolkowitz}}, \ and\
  \bibinfo {author} {\bibfnamefont {Alex}\ \bibnamefont {Levchenko}},\ }\href
  {https://arxiv.org/abs/2607.24906} {\enquote {\bibinfo {title} {Probing
  nonlocal superconducting fluctuations with covariance noise magnetometry},}\
  } (\bibinfo {year} {2026}),\ \Eprint {http://arxiv.org/abs/2607.24906}
  {arXiv:2607.24906 [cond-mat.supr-con]} \BibitemShut {NoStop}%
\bibitem [{\citenamefont {Ye}\ \emph {et~al.}(2012)\citenamefont {Ye},
  \citenamefont {Zhang}, \citenamefont {Akashi}, \citenamefont {Bahramy},
  \citenamefont {Arita},\ and\ \citenamefont {Iwasa}}]{Ye:2012}%
  \BibitemOpen
  \bibfield  {author} {\bibinfo {author} {\bibfnamefont {J.~T.}\ \bibnamefont
  {Ye}}, \bibinfo {author} {\bibfnamefont {Y.~J.}\ \bibnamefont {Zhang}},
  \bibinfo {author} {\bibfnamefont {R.}~\bibnamefont {Akashi}}, \bibinfo
  {author} {\bibfnamefont {M.~S.}\ \bibnamefont {Bahramy}}, \bibinfo {author}
  {\bibfnamefont {R.}~\bibnamefont {Arita}}, \ and\ \bibinfo {author}
  {\bibfnamefont {Y.}~\bibnamefont {Iwasa}},\ }\bibfield  {title} {\enquote
  {\bibinfo {title} {Superconducting dome in a gate-tuned band insulator},}\
  }\href@noop {} {\bibfield  {journal} {\bibinfo  {journal} {Science}\ }\textbf
  {\bibinfo {volume} {338}},\ \bibinfo {pages} {1193} (\bibinfo {year}
  {2012})}\BibitemShut {NoStop}%
\bibitem [{\citenamefont {Lu}\ \emph {et~al.}(2015)\citenamefont {Lu},
  \citenamefont {Zheliuk}, \citenamefont {Leermakers}, \citenamefont {Yuan},
  \citenamefont {Zeitler}, \citenamefont {Law},\ and\ \citenamefont
  {Ye}}]{Lu:2015}%
  \BibitemOpen
  \bibfield  {author} {\bibinfo {author} {\bibfnamefont {J.~M.}\ \bibnamefont
  {Lu}}, \bibinfo {author} {\bibfnamefont {O.}~\bibnamefont {Zheliuk}},
  \bibinfo {author} {\bibfnamefont {I.}~\bibnamefont {Leermakers}}, \bibinfo
  {author} {\bibfnamefont {N.~F.~Q.}\ \bibnamefont {Yuan}}, \bibinfo {author}
  {\bibfnamefont {U.}~\bibnamefont {Zeitler}}, \bibinfo {author} {\bibfnamefont
  {K.~T.}\ \bibnamefont {Law}}, \ and\ \bibinfo {author} {\bibfnamefont
  {J.~T.}\ \bibnamefont {Ye}},\ }\bibfield  {title} {\enquote {\bibinfo {title}
  {Evidence for two-dimensional {Ising} superconductivity in gated
  {MoS$_2$}},}\ }\href@noop {} {\bibfield  {journal} {\bibinfo  {journal}
  {Science}\ }\textbf {\bibinfo {volume} {350}},\ \bibinfo {pages} {1353}
  (\bibinfo {year} {2015})}\BibitemShut {NoStop}%
\bibitem [{\citenamefont {Saito}\ \emph {et~al.}(2016)\citenamefont {Saito},
  \citenamefont {Nakamura}, \citenamefont {Bahramy}, \citenamefont {Kohama},
  \citenamefont {Ye}, \citenamefont {Kasahara}, \citenamefont {Nakagawa},
  \citenamefont {Onga}, \citenamefont {Tokunaga}, \citenamefont {Nojima},
  \citenamefont {Yanase},\ and\ \citenamefont {Iwasa}}]{Saito:2016}%
  \BibitemOpen
  \bibfield  {author} {\bibinfo {author} {\bibfnamefont {Y.}~\bibnamefont
  {Saito}}, \bibinfo {author} {\bibfnamefont {Y.}~\bibnamefont {Nakamura}},
  \bibinfo {author} {\bibfnamefont {M.~S.}\ \bibnamefont {Bahramy}}, \bibinfo
  {author} {\bibfnamefont {Y.}~\bibnamefont {Kohama}}, \bibinfo {author}
  {\bibfnamefont {J.}~\bibnamefont {Ye}}, \bibinfo {author} {\bibfnamefont
  {Y.}~\bibnamefont {Kasahara}}, \bibinfo {author} {\bibfnamefont
  {Y.}~\bibnamefont {Nakagawa}}, \bibinfo {author} {\bibfnamefont
  {M.}~\bibnamefont {Onga}}, \bibinfo {author} {\bibfnamefont {M.}~\bibnamefont
  {Tokunaga}}, \bibinfo {author} {\bibfnamefont {T.}~\bibnamefont {Nojima}},
  \bibinfo {author} {\bibfnamefont {Y.}~\bibnamefont {Yanase}}, \ and\ \bibinfo
  {author} {\bibfnamefont {Y.}~\bibnamefont {Iwasa}},\ }\bibfield  {title}
  {\enquote {\bibinfo {title} {Superconductivity protected by spin-valley
  locking in ion-gated {MoS$_2$}},}\ }\href@noop {} {\bibfield  {journal}
  {\bibinfo  {journal} {Nat. Phys.}\ }\textbf {\bibinfo {volume} {12}},\
  \bibinfo {pages} {144} (\bibinfo {year} {2016})}\BibitemShut {NoStop}%
\bibitem [{\citenamefont {Xi}\ \emph {et~al.}(2016)\citenamefont {Xi},
  \citenamefont {Wang}, \citenamefont {Zhao}, \citenamefont {Park},
  \citenamefont {Law}, \citenamefont {Berger}, \citenamefont {Forr{\'o}},
  \citenamefont {Shan},\ and\ \citenamefont {Mak}}]{Xi:2016}%
  \BibitemOpen
  \bibfield  {author} {\bibinfo {author} {\bibfnamefont {X.}~\bibnamefont
  {Xi}}, \bibinfo {author} {\bibfnamefont {Z.}~\bibnamefont {Wang}}, \bibinfo
  {author} {\bibfnamefont {W.}~\bibnamefont {Zhao}}, \bibinfo {author}
  {\bibfnamefont {J.-H.}\ \bibnamefont {Park}}, \bibinfo {author}
  {\bibfnamefont {K.~T.}\ \bibnamefont {Law}}, \bibinfo {author} {\bibfnamefont
  {H.}~\bibnamefont {Berger}}, \bibinfo {author} {\bibfnamefont
  {L.}~\bibnamefont {Forr{\'o}}}, \bibinfo {author} {\bibfnamefont
  {J.}~\bibnamefont {Shan}}, \ and\ \bibinfo {author} {\bibfnamefont {K.~F.}\
  \bibnamefont {Mak}},\ }\bibfield  {title} {\enquote {\bibinfo {title} {Ising
  pairing in superconducting {NbSe$_2$} atomic layers},}\ }\href@noop {}
  {\bibfield  {journal} {\bibinfo  {journal} {Nat. Phys.}\ }\textbf {\bibinfo
  {volume} {12}},\ \bibinfo {pages} {139} (\bibinfo {year} {2016})}\BibitemShut
  {NoStop}%
\bibitem [{\citenamefont {Itahashi}\ \emph {et~al.}(2020)\citenamefont
  {Itahashi}, \citenamefont {Ideue}, \citenamefont {Saito}, \citenamefont
  {Shimizu}, \citenamefont {Ouchi}, \citenamefont {Nojima},\ and\ \citenamefont
  {Iwasa}}]{Itahashi:2020}%
  \BibitemOpen
  \bibfield  {author} {\bibinfo {author} {\bibfnamefont {Y.~M.}\ \bibnamefont
  {Itahashi}}, \bibinfo {author} {\bibfnamefont {T.}~\bibnamefont {Ideue}},
  \bibinfo {author} {\bibfnamefont {Y.}~\bibnamefont {Saito}}, \bibinfo
  {author} {\bibfnamefont {S.}~\bibnamefont {Shimizu}}, \bibinfo {author}
  {\bibfnamefont {T.}~\bibnamefont {Ouchi}}, \bibinfo {author} {\bibfnamefont
  {T.}~\bibnamefont {Nojima}}, \ and\ \bibinfo {author} {\bibfnamefont
  {Y.}~\bibnamefont {Iwasa}},\ }\bibfield  {title} {\enquote {\bibinfo {title}
  {Nonreciprocal transport in gate-induced polar superconductor {SrTiO$_3$}},}\
  }\href@noop {} {\bibfield  {journal} {\bibinfo  {journal} {Sci. Adv.}\
  }\textbf {\bibinfo {volume} {6}},\ \bibinfo {pages} {eaay9120} (\bibinfo
  {year} {2020})}\BibitemShut {NoStop}%
\bibitem [{\citenamefont {Rikken}\ \emph {et~al.}(2001)\citenamefont {Rikken},
  \citenamefont {F{\"o}lling},\ and\ \citenamefont {Wyder}}]{Rikken:2001}%
  \BibitemOpen
  \bibfield  {author} {\bibinfo {author} {\bibfnamefont {G.~L. J.~A.}\
  \bibnamefont {Rikken}}, \bibinfo {author} {\bibfnamefont {J.}~\bibnamefont
  {F{\"o}lling}}, \ and\ \bibinfo {author} {\bibfnamefont {P.}~\bibnamefont
  {Wyder}},\ }\bibfield  {title} {\enquote {\bibinfo {title} {Electrical
  magnetochiral anisotropy},}\ }\href@noop {} {\bibfield  {journal} {\bibinfo
  {journal} {Phys. Rev. Lett.}\ }\textbf {\bibinfo {volume} {87}},\ \bibinfo
  {pages} {236602} (\bibinfo {year} {2001})}\BibitemShut {NoStop}%
\bibitem [{\citenamefont {Wakatsuki}\ \emph {et~al.}(2017)\citenamefont
  {Wakatsuki}, \citenamefont {Saito}, \citenamefont {Hoshino}, \citenamefont
  {Itahashi}, \citenamefont {Ideue}, \citenamefont {Ezawa}, \citenamefont
  {Iwasa},\ and\ \citenamefont {Nagaosa}}]{Wakatsuki:2017}%
  \BibitemOpen
  \bibfield  {author} {\bibinfo {author} {\bibfnamefont {R.}~\bibnamefont
  {Wakatsuki}}, \bibinfo {author} {\bibfnamefont {Y.}~\bibnamefont {Saito}},
  \bibinfo {author} {\bibfnamefont {S.}~\bibnamefont {Hoshino}}, \bibinfo
  {author} {\bibfnamefont {Y.~M.}\ \bibnamefont {Itahashi}}, \bibinfo {author}
  {\bibfnamefont {T.}~\bibnamefont {Ideue}}, \bibinfo {author} {\bibfnamefont
  {M.}~\bibnamefont {Ezawa}}, \bibinfo {author} {\bibfnamefont
  {Y.}~\bibnamefont {Iwasa}}, \ and\ \bibinfo {author} {\bibfnamefont
  {N.}~\bibnamefont {Nagaosa}},\ }\bibfield  {title} {\enquote {\bibinfo
  {title} {Nonreciprocal charge transport in noncentrosymmetric
  superconductors},}\ }\href@noop {} {\bibfield  {journal} {\bibinfo  {journal}
  {Sci. Adv.}\ }\textbf {\bibinfo {volume} {3}},\ \bibinfo {pages} {e1602390}
  (\bibinfo {year} {2017})}\BibitemShut {NoStop}%
\bibitem [{\citenamefont {Tokura}\ and\ \citenamefont
  {Nagaosa}(2018)}]{TokuraNagaosa:2018}%
  \BibitemOpen
  \bibfield  {author} {\bibinfo {author} {\bibfnamefont {Y.}~\bibnamefont
  {Tokura}}\ and\ \bibinfo {author} {\bibfnamefont {N.}~\bibnamefont
  {Nagaosa}},\ }\bibfield  {title} {\enquote {\bibinfo {title} {Nonreciprocal
  responses from non-centrosymmetric quantum materials},}\ }\href@noop {}
  {\bibfield  {journal} {\bibinfo  {journal} {Nat. Commun.}\ }\textbf {\bibinfo
  {volume} {9}},\ \bibinfo {pages} {3740} (\bibinfo {year} {2018})}\BibitemShut
  {NoStop}%
\bibitem [{\citenamefont {Ando}\ \emph {et~al.}(2020)\citenamefont {Ando},
  \citenamefont {Miyasaka}, \citenamefont {Li}, \citenamefont {Ishizuka},
  \citenamefont {Arakawa}, \citenamefont {Shiota}, \citenamefont {Moriyama},
  \citenamefont {Yanase},\ and\ \citenamefont {Ono}}]{Ando:2020}%
  \BibitemOpen
  \bibfield  {author} {\bibinfo {author} {\bibfnamefont {F.}~\bibnamefont
  {Ando}}, \bibinfo {author} {\bibfnamefont {Y.}~\bibnamefont {Miyasaka}},
  \bibinfo {author} {\bibfnamefont {T.}~\bibnamefont {Li}}, \bibinfo {author}
  {\bibfnamefont {J.}~\bibnamefont {Ishizuka}}, \bibinfo {author}
  {\bibfnamefont {T.}~\bibnamefont {Arakawa}}, \bibinfo {author} {\bibfnamefont
  {Y.}~\bibnamefont {Shiota}}, \bibinfo {author} {\bibfnamefont
  {T.}~\bibnamefont {Moriyama}}, \bibinfo {author} {\bibfnamefont
  {Y.}~\bibnamefont {Yanase}}, \ and\ \bibinfo {author} {\bibfnamefont
  {T.}~\bibnamefont {Ono}},\ }\bibfield  {title} {\enquote {\bibinfo {title}
  {Observation of superconducting diode effect},}\ }\href@noop {} {\bibfield
  {journal} {\bibinfo  {journal} {Nature}\ }\textbf {\bibinfo {volume} {584}},\
  \bibinfo {pages} {373} (\bibinfo {year} {2020})}\BibitemShut {NoStop}%
\bibitem [{\citenamefont {Nadeem}\ \emph {et~al.}(2023)\citenamefont {Nadeem},
  \citenamefont {Fuhrer},\ and\ \citenamefont {Wang}}]{Nadeem:2023}%
  \BibitemOpen
  \bibfield  {author} {\bibinfo {author} {\bibfnamefont {M.}~\bibnamefont
  {Nadeem}}, \bibinfo {author} {\bibfnamefont {M.~S.}\ \bibnamefont {Fuhrer}},
  \ and\ \bibinfo {author} {\bibfnamefont {X.}~\bibnamefont {Wang}},\
  }\bibfield  {title} {\enquote {\bibinfo {title} {The superconducting diode
  effect},}\ }\href@noop {} {\bibfield  {journal} {\bibinfo  {journal} {Nat.
  Rev. Phys.}\ }\textbf {\bibinfo {volume} {5}},\ \bibinfo {pages} {558}
  (\bibinfo {year} {2023})}\BibitemShut {NoStop}%
\bibitem [{\citenamefont {Shaffer}\ and\ \citenamefont
  {Levchenko}(2025)}]{Shaffer:2025}%
  \BibitemOpen
  \bibfield  {author} {\bibinfo {author} {\bibfnamefont {Daniel}\ \bibnamefont
  {Shaffer}}\ and\ \bibinfo {author} {\bibfnamefont {Alex}\ \bibnamefont
  {Levchenko}},\ }\href {https://arxiv.org/abs/2510.25864} {\enquote {\bibinfo
  {title} {Theories of superconducting diode effects},}\ } (\bibinfo {year}
  {2025}),\ \Eprint {http://arxiv.org/abs/2510.25864} {arXiv:2510.25864
  [cond-mat.supr-con]} \BibitemShut {NoStop}%
\bibitem [{\citenamefont {Feigel'man}\ \emph {et~al.}(2000)\citenamefont
  {Feigel'man}, \citenamefont {Larkin},\ and\ \citenamefont
  {Skvortsov}}]{FLS:2000}%
  \BibitemOpen
  \bibfield  {author} {\bibinfo {author} {\bibfnamefont {M.~V.}\ \bibnamefont
  {Feigel'man}}, \bibinfo {author} {\bibfnamefont {A.~I.}\ \bibnamefont
  {Larkin}}, \ and\ \bibinfo {author} {\bibfnamefont {M.~A.}\ \bibnamefont
  {Skvortsov}},\ }\bibfield  {title} {\enquote {\bibinfo {title} {Keldysh
  action for disordered superconductors},}\ }\href@noop {} {\bibfield
  {journal} {\bibinfo  {journal} {Phys. Rev. B}\ }\textbf {\bibinfo {volume}
  {61}},\ \bibinfo {pages} {12361} (\bibinfo {year} {2000})}\BibitemShut
  {NoStop}%
\bibitem [{\citenamefont {Yurkevich}\ and\ \citenamefont
  {Lerner}(2001)}]{Yurkevich:2001}%
  \BibitemOpen
  \bibfield  {author} {\bibinfo {author} {\bibfnamefont {I.~V.}\ \bibnamefont
  {Yurkevich}}\ and\ \bibinfo {author} {\bibfnamefont {Igor~V.}\ \bibnamefont
  {Lerner}},\ }\bibfield  {title} {\enquote {\bibinfo {title} {Nonlinear
  $\ensuremath{\sigma}$ model for disordered superconductors},}\ }\href
  {\doibase 10.1103/PhysRevB.63.064522} {\bibfield  {journal} {\bibinfo
  {journal} {Phys. Rev. B}\ }\textbf {\bibinfo {volume} {63}},\ \bibinfo
  {pages} {064522} (\bibinfo {year} {2001})}\BibitemShut {NoStop}%
\bibitem [{\citenamefont {Levchenko}\ and\ \citenamefont
  {Kamenev}(2007)}]{LevchenkoKamenev:2007}%
  \BibitemOpen
  \bibfield  {author} {\bibinfo {author} {\bibfnamefont {A.}~\bibnamefont
  {Levchenko}}\ and\ \bibinfo {author} {\bibfnamefont {A.}~\bibnamefont
  {Kamenev}},\ }\bibfield  {title} {\enquote {\bibinfo {title} {Keldysh
  {Ginzburg-Landau} action of fluctuating superconductors},}\ }\href@noop {}
  {\bibfield  {journal} {\bibinfo  {journal} {Phys. Rev. B}\ }\textbf {\bibinfo
  {volume} {76}},\ \bibinfo {pages} {094518} (\bibinfo {year}
  {2007})}\BibitemShut {NoStop}%
\bibitem [{\citenamefont {Kamenev}(2011)}]{KamenevBook}%
  \BibitemOpen
  \bibfield  {author} {\bibinfo {author} {\bibfnamefont {A.}~\bibnamefont
  {Kamenev}},\ }\href@noop {} {\emph {\bibinfo {title} {Field Theory of
  Non-Equilibrium Systems}}}\ (\bibinfo  {publisher} {Cambridge University
  Press},\ \bibinfo {address} {Cambridge},\ \bibinfo {year} {2011})\BibitemShut
  {NoStop}%
\bibitem [{\citenamefont {Liao}\ \emph {et~al.}(2017)\citenamefont {Liao},
  \citenamefont {Levchenko},\ and\ \citenamefont {Foster}}]{Liao:2017}%
  \BibitemOpen
  \bibfield  {author} {\bibinfo {author} {\bibfnamefont {Yunxiang}\
  \bibnamefont {Liao}}, \bibinfo {author} {\bibfnamefont {Alex}\ \bibnamefont
  {Levchenko}}, \ and\ \bibinfo {author} {\bibfnamefont {Matthew~S.}\
  \bibnamefont {Foster}},\ }\bibfield  {title} {\enquote {\bibinfo {title}
  {Response theory of the ergodic many-body delocalized phase: {Keldysh
  Finkel'stein} sigma models and the 10-fold way},}\ }\href@noop {} {\bibfield
  {journal} {\bibinfo  {journal} {Annals of Physics}\ }\textbf {\bibinfo
  {volume} {386}},\ \bibinfo {pages} {97--157} (\bibinfo {year}
  {2017})}\BibitemShut {NoStop}%
\bibitem [{\citenamefont {Virtanen}\ \emph {et~al.}(2021)\citenamefont
  {Virtanen}, \citenamefont {Bergeret},\ and\ \citenamefont
  {Tokatly}}]{VBT:2021}%
  \BibitemOpen
  \bibfield  {author} {\bibinfo {author} {\bibfnamefont {P.}~\bibnamefont
  {Virtanen}}, \bibinfo {author} {\bibfnamefont {F.~S.}\ \bibnamefont
  {Bergeret}}, \ and\ \bibinfo {author} {\bibfnamefont {I.~V.}\ \bibnamefont
  {Tokatly}},\ }\bibfield  {title} {\enquote {\bibinfo {title} {Magnetoelectric
  effects in superconductors due to spin-orbit scattering: Nonlinear
  $\sigma$-model description},}\ }\href@noop {} {\bibfield  {journal} {\bibinfo
   {journal} {Phys. Rev. B}\ }\textbf {\bibinfo {volume} {104}},\ \bibinfo
  {pages} {064515} (\bibinfo {year} {2021})}\BibitemShut {NoStop}%
\bibitem [{\citenamefont {Virtanen}\ \emph {et~al.}(2022)\citenamefont
  {Virtanen}, \citenamefont {Bergeret},\ and\ \citenamefont
  {Tokatly}}]{VBT:2022}%
  \BibitemOpen
  \bibfield  {author} {\bibinfo {author} {\bibfnamefont {P.}~\bibnamefont
  {Virtanen}}, \bibinfo {author} {\bibfnamefont {F.~S.}\ \bibnamefont
  {Bergeret}}, \ and\ \bibinfo {author} {\bibfnamefont {I.~V.}\ \bibnamefont
  {Tokatly}},\ }\bibfield  {title} {\enquote {\bibinfo {title} {Nonlinear
  $\sigma$ model for disordered systems with intrinsic spin-orbit coupling},}\
  }\href@noop {} {\bibfield  {journal} {\bibinfo  {journal} {Phys. Rev. B}\
  }\textbf {\bibinfo {volume} {105}},\ \bibinfo {pages} {224517} (\bibinfo
  {year} {2022})}\BibitemShut {NoStop}%
\bibitem [{\citenamefont {Bergeret}\ and\ \citenamefont
  {Tokatly}(2014)}]{BergeretTokatly:2014}%
  \BibitemOpen
  \bibfield  {author} {\bibinfo {author} {\bibfnamefont {F.~S.}\ \bibnamefont
  {Bergeret}}\ and\ \bibinfo {author} {\bibfnamefont {I.~V.}\ \bibnamefont
  {Tokatly}},\ }\bibfield  {title} {\enquote {\bibinfo {title} {Spin-orbit
  coupling as a source of long-range triplet proximity effect in
  superconductor-ferromagnet hybrid structures},}\ }\href@noop {} {\bibfield
  {journal} {\bibinfo  {journal} {Phys. Rev. B}\ }\textbf {\bibinfo {volume}
  {89}},\ \bibinfo {pages} {134517} (\bibinfo {year} {2014})}\BibitemShut
  {NoStop}%
\bibitem [{\citenamefont {Liu}\ \emph {et~al.}(2026)\citenamefont {Liu},
  \citenamefont {de~Miranda}, \citenamefont {Shaffer},\ and\ \citenamefont
  {Levchenko}}]{Liu:2026}%
  \BibitemOpen
  \bibfield  {author} {\bibinfo {author} {\bibfnamefont {Tony}\ \bibnamefont
  {Liu}}, \bibinfo {author} {\bibfnamefont {Joaquim~Telles}\ \bibnamefont
  {de~Miranda}}, \bibinfo {author} {\bibfnamefont {Daniel}\ \bibnamefont
  {Shaffer}}, \ and\ \bibinfo {author} {\bibfnamefont {Alex}\ \bibnamefont
  {Levchenko}},\ }\href {https://arxiv.org/abs/2608.05306} {\enquote {\bibinfo
  {title} {Kinetic {Lifshitz} invariants and dynamics of nonreciprocal
  fluctuations in superconductors},}\ } (\bibinfo {year} {2026}),\ \Eprint
  {http://arxiv.org/abs/2608.05306} {arXiv:2608.05306 [cond-mat.supr-con]}
  \BibitemShut {NoStop}%
\bibitem [{\citenamefont {Mineev}\ and\ \citenamefont
  {Samokhin}(1994)}]{MineevSamokhin:1994}%
  \BibitemOpen
  \bibfield  {author} {\bibinfo {author} {\bibfnamefont {V.~P.}\ \bibnamefont
  {Mineev}}\ and\ \bibinfo {author} {\bibfnamefont {K.~V.}\ \bibnamefont
  {Samokhin}},\ }\bibfield  {title} {\enquote {\bibinfo {title} {Helical phases
  in superconductors},}\ }\href@noop {} {\bibfield  {journal} {\bibinfo
  {journal} {Zh. Eksp. Teor. Fiz.}\ }\textbf {\bibinfo {volume} {105}},\
  \bibinfo {pages} {747} (\bibinfo {year} {1994})},\ \bibinfo {note} {[JETP
  {\bf 78}, 401 (1994)]}\BibitemShut {NoStop}%
\bibitem [{\citenamefont {Dimitrova}\ and\ \citenamefont
  {Feigel'man}(2007)}]{DimitrovaFeigelman:2007}%
  \BibitemOpen
  \bibfield  {author} {\bibinfo {author} {\bibfnamefont {O.}~\bibnamefont
  {Dimitrova}}\ and\ \bibinfo {author} {\bibfnamefont {M.~V.}\ \bibnamefont
  {Feigel'man}},\ }\bibfield  {title} {\enquote {\bibinfo {title} {Theory of a
  two-dimensional superconductor with broken inversion symmetry},}\ }\href@noop
  {} {\bibfield  {journal} {\bibinfo  {journal} {Phys. Rev. B}\ }\textbf
  {\bibinfo {volume} {76}},\ \bibinfo {pages} {014522} (\bibinfo {year}
  {2007})}\BibitemShut {NoStop}%
\bibitem [{\citenamefont {Kaur}\ \emph {et~al.}(2005)\citenamefont {Kaur},
  \citenamefont {Agterberg},\ and\ \citenamefont {Sigrist}}]{Kaur:2005}%
  \BibitemOpen
  \bibfield  {author} {\bibinfo {author} {\bibfnamefont {R.~P.}\ \bibnamefont
  {Kaur}}, \bibinfo {author} {\bibfnamefont {D.~F.}\ \bibnamefont {Agterberg}},
  \ and\ \bibinfo {author} {\bibfnamefont {M.}~\bibnamefont {Sigrist}},\
  }\bibfield  {title} {\enquote {\bibinfo {title} {Helical vortex phase in the
  noncentrosymmetric {CePt$_3$Si}},}\ }\href@noop {} {\bibfield  {journal}
  {\bibinfo  {journal} {Phys. Rev. Lett.}\ }\textbf {\bibinfo {volume} {94}},\
  \bibinfo {pages} {137002} (\bibinfo {year} {2005})}\BibitemShut {NoStop}%
\bibitem [{\citenamefont {Edelstein}(1995)}]{Edelstein:1995}%
  \BibitemOpen
  \bibfield  {author} {\bibinfo {author} {\bibfnamefont {V.~M.}\ \bibnamefont
  {Edelstein}},\ }\bibfield  {title} {\enquote {\bibinfo {title}
  {Magnetoelectric effect in polar superconductors},}\ }\href@noop {}
  {\bibfield  {journal} {\bibinfo  {journal} {Phys. Rev. Lett.}\ }\textbf
  {\bibinfo {volume} {75}},\ \bibinfo {pages} {2004} (\bibinfo {year}
  {1995})}\BibitemShut {NoStop}%
\bibitem [{\citenamefont {Ili{\'c}}\ and\ \citenamefont
  {Bergeret}(2022)}]{IlicBergeret:2022}%
  \BibitemOpen
  \bibfield  {author} {\bibinfo {author} {\bibfnamefont {S.}~\bibnamefont
  {Ili{\'c}}}\ and\ \bibinfo {author} {\bibfnamefont {F.~S.}\ \bibnamefont
  {Bergeret}},\ }\bibfield  {title} {\enquote {\bibinfo {title} {Theory of the
  supercurrent diode effect in {Rashba} superconductors with arbitrary
  disorder},}\ }\href@noop {} {\bibfield  {journal} {\bibinfo  {journal} {Phys.
  Rev. Lett.}\ }\textbf {\bibinfo {volume} {128}},\ \bibinfo {pages} {177001}
  (\bibinfo {year} {2022})}\BibitemShut {NoStop}%
\bibitem [{\citenamefont {Hasan}\ \emph {et~al.}(2024)\citenamefont {Hasan},
  \citenamefont {Shaffer}, \citenamefont {Khodas},\ and\ \citenamefont
  {Levchenko}}]{Hasan:2024}%
  \BibitemOpen
  \bibfield  {author} {\bibinfo {author} {\bibfnamefont {Jaglul}\ \bibnamefont
  {Hasan}}, \bibinfo {author} {\bibfnamefont {Daniel}\ \bibnamefont {Shaffer}},
  \bibinfo {author} {\bibfnamefont {Maxim}\ \bibnamefont {Khodas}}, \ and\
  \bibinfo {author} {\bibfnamefont {Alex}\ \bibnamefont {Levchenko}},\
  }\bibfield  {title} {\enquote {\bibinfo {title} {Supercurrent diode effect in
  helical superconductors},}\ }\href {\doibase 10.1103/PhysRevB.110.024508}
  {\bibfield  {journal} {\bibinfo  {journal} {Phys. Rev. B}\ }\textbf {\bibinfo
  {volume} {110}},\ \bibinfo {pages} {024508} (\bibinfo {year}
  {2024})}\BibitemShut {NoStop}%
\bibitem [{\citenamefont {Hasan}\ \emph {et~al.}(2025)\citenamefont {Hasan},
  \citenamefont {Shaffer}, \citenamefont {Khodas},\ and\ \citenamefont
  {Levchenko}}]{Hasan:2025}%
  \BibitemOpen
  \bibfield  {author} {\bibinfo {author} {\bibfnamefont {Jaglul}\ \bibnamefont
  {Hasan}}, \bibinfo {author} {\bibfnamefont {Daniel}\ \bibnamefont {Shaffer}},
  \bibinfo {author} {\bibfnamefont {Maxim}\ \bibnamefont {Khodas}}, \ and\
  \bibinfo {author} {\bibfnamefont {Alex}\ \bibnamefont {Levchenko}},\
  }\bibfield  {title} {\enquote {\bibinfo {title} {Superconducting diode
  efficiency from singlet-triplet mixing in disordered systems},}\ }\href
  {\doibase 10.1103/PhysRevB.111.174514} {\bibfield  {journal} {\bibinfo
  {journal} {Phys. Rev. B}\ }\textbf {\bibinfo {volume} {111}},\ \bibinfo
  {pages} {174514} (\bibinfo {year} {2025})}\BibitemShut {NoStop}%
\bibitem [{\citenamefont {Cameron}\ and\ \citenamefont
  {Martin}(1945)}]{CameronMartin:1945}%
  \BibitemOpen
  \bibfield  {author} {\bibinfo {author} {\bibfnamefont {R.~H.}\ \bibnamefont
  {Cameron}}\ and\ \bibinfo {author} {\bibfnamefont {W.~T.}\ \bibnamefont
  {Martin}},\ }\bibfield  {title} {\enquote {\bibinfo {title} {Evaluations of
  various {Wiener} integrals by use of certain {Sturm--Liouville} differential
  equations},}\ }\href@noop {} {\bibfield  {journal} {\bibinfo  {journal}
  {Bull. Am. Math. Soc.}\ }\textbf {\bibinfo {volume} {51}},\ \bibinfo {pages}
  {73--90} (\bibinfo {year} {1945})}\BibitemShut {NoStop}%
\bibitem [{\citenamefont {Jakeman}\ and\ \citenamefont
  {Pike}(1968)}]{JakemanPike:1968}%
  \BibitemOpen
  \bibfield  {author} {\bibinfo {author} {\bibfnamefont {E.}~\bibnamefont
  {Jakeman}}\ and\ \bibinfo {author} {\bibfnamefont {E.~R.}\ \bibnamefont
  {Pike}},\ }\bibfield  {title} {\enquote {\bibinfo {title} {The
  intensity-fluctuation distribution of {Gaussian} light},}\ }\href@noop {}
  {\bibfield  {journal} {\bibinfo  {journal} {J. Phys. A: Gen. Phys.}\ }\textbf
  {\bibinfo {volume} {1}},\ \bibinfo {pages} {128--138} (\bibinfo {year}
  {1968})}\BibitemShut {NoStop}%
\bibitem [{\citenamefont {de~Miranda}\ and\ \citenamefont
  {Levchenko}(2026)}]{JTM:2026}%
  \BibitemOpen
  \bibfield  {author} {\bibinfo {author} {\bibfnamefont {Joaquim~Telles}\
  \bibnamefont {de~Miranda}}\ and\ \bibinfo {author} {\bibfnamefont {Alex}\
  \bibnamefont {Levchenko}},\ }\href {https://arxiv.org/abs/2608.20166}
  {\enquote {\bibinfo {title} {Photogalvanic transport of nonreciprocal
  {Cooper}-pair fluctuations},}\ } (\bibinfo {year} {2026}),\ \Eprint
  {http://arxiv.org/abs/2608.20166} {arXiv:2608.20166 [cond-mat.supr-con]}
  \BibitemShut {NoStop}%
\end{thebibliography}%

\begin{center}
\textbf{\large END MATTER}
\end{center}
\vspace{0.5em}

\section*{S1. Fluctuation propagators from the sigma model}

The propagators (\ref{eq:LR})--(\ref{eq:LK}) summarize the Gaussian sector of the Keldysh NLSM effective action of a disordered Rashba film in an in-plane field, derived in the companion work \cite{Liu:2026}. The singlet Cooperon is coupled to the triplet Cooperons by the Zeeman vertex ($-2ih$) and by the SU(2) field-strength vertex ($4i\kappa_0q$, $\kappa_0=\tfrac{\alpha}{2}(\alpha p_F\tau)^2$). The product of the two vertices gives all couplings odd in momentum. The intermediate triplet propagator is limited by the temperature or by the Dyakonov-Perel rate $\Gamma_t=2(\alpha p_F)^2\tau$, whichever is larger. This leads to a single crossover function of $\Gamma_t/T$ for the coefficients of the Gaussian sector. In the weak-relaxation regime $\Gamma_t\ll4\pi T$ used in the main text the parameters $u$, $\Xi$, and $q_0$ take the values quoted in Eq.~(\ref{eq:params}). In the opposite limit $q_0\to4\alpha h/v_F^2$ does not depend on disorder. The odd couplings then reduce, to first order in $h$, to a shift of the pair momentum, and field-odd observables of the film contain the factor $R\simeq(2/\pi^2)(\ln g)/g$, $g=\Gamma_t/4\pi T$, multiplying $2u$ at the given $\Gamma_t/T$. At weak relaxation $R=0.41$ \cite{Liu:2026}. The equilibrium relation $\mathcal L_K^{-1}=\coth(\Omega/2T)[\mathcal L_R^{-1}-\mathcal L_A^{-1}]$ is exact and holds with all magnetoelectric terms included. For this reason the asymmetry of the noise is equal to the asymmetry of the friction. In real space the Langevin correlator is $\langle\xi_\Delta(\bm r)\xi^*_\Delta(\bm r')\rangle=(16T^2/\pi\nu)[\delta(\bm r-\bm r')-i\bm u\cdot\nabla_{\bm r}\delta(\bm r-\bm r')]$ per unit time. This form refers to the TDGL equation written with unit coefficient of $\partial_t\Delta$, $(1+\bm u\cdot\bm q)\partial_t\Delta=-(8T/\pi)a(\bm q)\Delta+\xi_\Delta$, as in Ref.~\cite{LevchenkoKamenev:2007}. If the kinetic coefficient $(\pi/8T)(1+\bm u\cdot\bm q)$ is kept in front of $\partial_t\Delta$, as in Ref.~\cite{Liu:2026}, the intensity is $(\pi/4\nu)(1+\bm u\cdot\bm q)$.

\section*{S2. Junction noise: frequency convolution and the mismatch function}

\emph{Mismatch function.}  Squaring the fluctuational Josephson current and pairing with Eq.~(\ref{eq:LK}) for each electrode gives $S_J(\omega)=\sum_\pm C_J\int\!\frac{d^2q}{(2\pi)^2}\int\!\frac{d\Omega}{2\pi}\,\mathcal L_K^L(\bm q,\Omega)\mathcal L_K^R(\bm q,\Omega-\omega_\pm)$. Each factor is a Lorentzian $\ell_a(\Omega)=(1+w_a)/[\varepsilon_a^2+(1+w_a)^2\Omega^2]$ of width $\bar\varepsilon_a=\varepsilon_a/(1+w_a)$, with $w_a=\bm u_a\cdot\bm q$; the frequency integral is
\begin{equation}
\int\!\frac{d\Omega}{2\pi}\,
\ell_L(\Omega)\,\ell_R(\Omega-s)
=\frac{(\bar\varepsilon_L+\bar\varepsilon_R)(1+w_L)^{-1}(1+w_R)^{-1}}
{2\bar\varepsilon_L\bar\varepsilon_R\big[(\bar\varepsilon_L+\bar\varepsilon_R)^2+s^2\big]}.
\end{equation}
An expansion of the prefactors shows that the vectors $\bm u_a$ enter $S_J$ only at second order. To leading order the nonreciprocal terms enter Eq.~(\ref{eq:SJ}) through $\varepsilon_{L,R}$. The change of variable $\bm q\to-\bm q$ together with $\mathcal L(\bm q;\bm h)=\mathcal L(-\bm q;-\bm h)$ proves the symmetry $S(\omega;\bm h)=S(\omega;-\bm h)$ for barriers that conserve the in-plane momentum. For a diffusive barrier the two momentum integrals are independent. The symmetry then holds as well, but the dependence on the mismatch $\kappa$ disappears. A barrier that partially conserves momentum retains this dependence. Removing the common shift $(\bm q_0^L+\bm q_0^R)/2$ and rescaling $\bm q=\bm k/\sqrt{D\tau_{\mathrm{GL}}}$ reduces the peak height to the function $P(\kappa)$ of Eq.~(\ref{eq:P}), with $\varepsilon_L+\varepsilon_R=2\tau_{\mathrm{GL}}^{-1}(1+k^2+\kappa^2)$. Expanding $[(1+(\bm k-\bm\kappa)^2)(1+(\bm k+\bm\kappa)^2)]^{-1}=[A^2-4k_x^2\kappa^2]^{-1}$ with $A=1+k^2+\kappa^2$ and using $\int d^2k\,(1+k^2)^{-m}=\pi/(m-1)$, $\int d^2k\,k^2(1+k^2)^{-5}=\pi/12$ yields $P=1-\tfrac53\kappa^2+O(\kappa^4)$; the large-$\kappa$ tail follows from the two disjoint regions $\bm k\approx\pm\bm\kappa$ where only one of the two propagators is large. For equal electrodes at $\kappa=0$ the momentum integral evaluates to $(\tau^2_{\mathrm{GL}}/16\pi D)N(z)$ with $N(z)=(4/z^2)\ln\sqrt{1+z^2/4}$, which coincides with Eq.~(36) of Ref.~\cite{Levchenko:2008} and fixes the normalization of Eq.~(\ref{eq:SJfinal}).

\emph{Orbital mismatch.} An in-plane field $\bm H$ threading the barrier gives the two electrodes vector potentials that differ by $Hd_{\rm eff}$, with $d_{\rm eff}$ the center-to-center distance of the two thin films, so that their gauge-invariant pair momenta differ by $\bm q_{\rm orb}=2eHd_{\rm eff}\,\hat{\bm z}\times\hat{\bm H}$, i.e., $q_{\rm orb}=2\pi Hd_{\rm eff}/\Phi_0\approx3\,\mu\mathrm{m}^{-1}\,(H/0.1\,\mathrm{T})(d_{\rm eff}/10\,\mathrm{nm})$, collinear with $\bm n=\bm h\times\bm\alpha$ and hence with $\delta\bm q_0$. The helical shift is bounded by its strong-relaxation value $q_0=4\alpha h/v_F^2=2(\alpha p_F/E_F)(h/v_F)$, which is $\sim0.02\,\mu\mathrm{m}^{-1}$ at $0.1\,$T for $\alpha p_F/E_F=0.1$, $v_F=10^5\,$m/s and $h=\mu_BH$, and is smaller by $\simeq0.14\,\Gamma_t/T$ in the weak-relaxation regime of Eq.~(\ref{eq:params}); the orbital mismatch is therefore typically the larger of the two. It is the same for parallel and antiparallel polar axes and for any disorder, whereas the helical mismatch changes sign when one polar axis is reversed (or, in gate-tunable films, when the sign of $\alpha$ is reversed); the in-plane field also produces a common orbital pair breaking $\propto D(eHd)^2$, even in $h$, which is absorbed in $\epsilon$. Since $\kappa$ is defined through half the difference of the pair momenta of the two electrodes, $P(\kappa)$ is to be evaluated with $\bm\kappa=(\bm q_{\rm orb}/2+\delta\bm q_0)\xi_{\mathrm{GL}}$. The helical contribution is the difference of the peak heights for the two antiparallel arrangements of the polar axes, $P_+-P_-=P(|q_{\rm orb}/2+q_0|\xi_{\mathrm{GL}})-P(|q_{\rm orb}/2-q_0|\xi_{\mathrm{GL}})\simeq2q_0\xi_{\mathrm{GL}}\,\partial_\kappa P(q_{\rm orb}\xi_{\mathrm{GL}}/2)$, which for small $\kappa$ reduces to the expression quoted in the main text. The symmetry $S(\omega;\bm h)=S(\omega;-\bm h)$ is not affected, since $\bm q_{\rm orb}$ reverses with $\bm H$ and $\mathcal L(\bm q;\bm h)=\mathcal L(-\bm q;-\bm h)$ continues to hold.

\section*{S3. Driven film: drive expansion, master integrals, and spectral shapes}

Writing the TDGL dynamics per mode as $d\Delta_{\bm q}/dt=-\lambda_{\bm q}\Delta_{\bm q}+\eta_{\bm q}$ with $\lambda_{\bm q}=(8T/\pi)a(\bm q)/(1+\bm u\cdot\bm q)$ and FDT-normalized noise, $\langle\eta_{\bm q}\eta^*_{\bm q}\rangle=2\lambda_{\bm q}\bar n_{\bm q}$, where $a(\bm q)=(\pi/8T)\,\varepsilon(\bm q)=\epsilon+(\pi D/8T)q^2-\Lambda q_n+(\pi\Xi/8T)\,q_nq^2$ is the GL kernel and $\Lambda=(\pi D/4T)\,q_0$, the Gaussian current noise is $S(\omega)=\int_{\bm q}J^2n_{\bm q}^2\,4\lambda_{\bm q}/(\omega^2+4\lambda_{\bm q}^2)$ with $J=2e\nu\,\partial_{q_n}a$. In a dc field the kinetic momentum of the mode of canonical momentum $\bm q$ drifts, $\bm k(t)=\bm q+2e\bm Et$, so that the mode is an Ornstein-Uhlenbeck process with the time-dependent rate $\lambda(\bm k(t))$. The linear dynamics gives exactly $\langle\Delta(\tau)\Delta^*(0)\rangle=n_{\bm q}\exp[-\!\int_0^\tau\lambda(\bm k(s))ds]$, with $n_{\bm q}$ the stationary occupation, and Gaussianity gives $\langle|\Delta(\tau)|^2|\Delta(0)|^2\rangle_c=|\langle\Delta(\tau)\Delta^*(0)\rangle|^2$. With the current vertex evaluated at the kinetic momentum, the current correlator for $\tau>0$ is
\begin{equation}
S(\tau)=\!\int_{\bm q}\!n_{\bm q}^2\,J(\bm q)J(\bm q+2e\bm E\tau)\,
e^{-2\!\int_0^\tau\!\lambda(\bm q+2e\bm Es)ds},
\label{eq:SMcorr}
\end{equation}
and $S(\omega)=2\int_0^\infty d\tau\cos(\omega\tau)S(\tau)$. To linear order in $E$ there are three terms. The first comes from the change of the stationary occupation, $n^{(1)}_{\bm q}=-(e\bm E\cdot\partial_{\bm q}\bar n)/\lambda_{\bm q}$, which follows from the kinetic equation of the main text. The other two come from the drift of the vertex and of the rate. They are odd in $E$ and even in $\tau$, and therefore contribute to the cosine transform:
\begin{align}
\delta S(\omega)=\!\int_{\bm q}\!\Big\{&2J^2\bar n\,n^{(1)}\frac{4\lambda}{\omega^2+4\lambda^2}
\nonumber\\[-1pt]
&+2e\bm E\!\cdot\!\partial_{\bm q}J\;J\bar n^2\,\frac{2(4\lambda^2-\omega^2)}{(4\lambda^2+\omega^2)^2}
\nonumber\\[-1pt]
&-2e\bm E\!\cdot\!\partial_{\bm q}\lambda\;J^2\bar n^2\,\frac{8\lambda(4\lambda^2-3\omega^2)}{(4\lambda^2+\omega^2)^3}\Big\}.
\label{eq:SMmaster}
\end{align}
At $\omega=0$ one integration by parts brings the first term to $e\bm E\cdot\!\int_{\bm q}\bar n^2\partial_{\bm q}(J^2/\lambda^2)$ and the sum of the last two to exactly one half of it, for arbitrary $J$, $\lambda$, $\bar n$ and in any dimension. The angular and radial integrals give the coefficients $u:\ +\tfrac{3}{16}$, $\Xi/D:\ +\tfrac{3}{16}$, $\Lambda:\ 0$ in units of $1/8\pi\epsilon^2$ [the three terms of Eq.~(\ref{eq:SMmaster}) contribute $\tfrac18\{1,\tfrac23,-\tfrac16\}$ to the $u$ column and $\tfrac18\{1,1,-\tfrac12\}$ to the $\Xi/D$ column, and each of them cancels separately in the $\Lambda$ column], from which Eqs.~(\ref{eq:dSw})--(\ref{eq:Nyquist}) follow with $u+\Xi/D=(\alpha p_F\tau)^2\alpha h/T^2$ and, once the cross terms of $\Lambda$ with the $q^4$ term of the spectrum and the $q^2$ term of the friction are included, with $u+\Xi/D\to R\,(u+\Xi/D)$; the sum of the three terms cancels in the $\Lambda$ column at all $\omega$, since Eq.~(\ref{eq:SMcorr}) is invariant under $\bm q\to\bm q+\bm q_0$. The fluctuation relation (\ref{eq:Nyquist}) follows from these expressions as an exact identity. The occupation term and the MCA current reduce to
\begin{subequations}
\begin{align}
&\delta S^{\mathrm{occ}}_{\mathrm{odd}}(0)\propto I_n=\!\int_{\bm q}\!\frac{(\partial_{q_n}a)^3(1+w)^2}{a^5},
\\
&2T\,\partial_E\delta j_{\mathrm{MCA}}\propto -I_m
=-\!\int_{\bm q}\!(1+w)\frac{\partial_{q_n}a}{a}\,
\partial_{q_n}\!\Big[\frac{(1+w)\,\partial_{q_n}a}{a^3}\Big],
\end{align}
\end{subequations}
with equal prefactors, and a single integration by parts of the middle term of $I_m$ [using $(1+w)^2(\partial_{q_n}a)(\partial^2_{q_n}a)=\tfrac12(1+w)^2\partial_{q_n}(\partial_{q_n}a)^2$] gives $I_m=-I_n$ identically. This holds for arbitrary dispersion $a(\bm q)$, arbitrary drift $w(\bm q)=\bm u\cdot\bm q$, in any dimension, and to all orders in the odd couplings. The drift terms add one half of $I_n$, whence $\delta S_{\mathrm{odd}}(0)=3T\partial_E\delta j_{\mathrm{MCA}}$. The relation relies only on the assumptions of the kinetic equation. These are local equilibrium of each mode at its instantaneous momentum, as required by the FDT, Markovian dynamics of the modes, and the thermodynamic form of the current vertex \cite{Liu:2026}. At finite $\omega$ the $u$ part of Eq.~(\ref{eq:SMmaster}) reduces to $F_u(\Omega)=64\int_1^\infty dy\,(y-1)[32y^5-16y^4+40\Omega^2y^3-48\Omega^2y^2-4\Omega^4y+\Omega^4]/[y\,(\Omega^2+4y^2)^4]$, i.e., the closed form (\ref{eq:Fu}); the $\Xi$ part is the analogous combination normalized to unity at $\Omega=0$, which evaluates to Eq.~(\ref{eq:FXi}), with the logarithmically enhanced tail quoted in the main text; the sum of the two is Eq.~(\ref{eq:Fsum}). The terms linear in $\Lambda$ cancel at all $\Omega$, because the helical shift can be removed by a change of the integration variable. The same expressions give the equilibrium noise $S(0)=2T\sigma_{AL}$ with $\sigma_{AL}=e^2/16\epsilon$ and, at first order in $E$, the AL conductivity. This checks the signs and normalizations.

\section*{S4. Counting statistics}

For a complex Ornstein-Uhlenbeck mode, $\langle\Delta(t)\Delta^*(0)\rangle=\bar n e^{-\lambda|t|}$, the Cameron-Martin formula \cite{CameronMartin:1945} for the generating functional of $\int J|\Delta|^2dt$ gives the per-mode rate $\mu(\eta)=\lambda[1-\sqrt{1-2i\eta J\bar n/\lambda}]$, whose expansion reproduces $c_1=J\bar n$, $c_2=J^2\bar n^2/\lambda$, $c_3=3J^3\bar n^3/\lambda^2$, and generally $c_k=(2k-3)!!\,J^k\bar n^k\lambda^{1-k}$; summing over modes yields Eq.~(\ref{eq:CGF}). The same formula describes the photocount statistics of thermal (Gaussian) light \cite{JakemanPike:1968}. The equilibrium current vanishes identically, $\sum_{\bm q}J\bar n\propto\int d^2q\,\partial_{q_n}a/a=0$ (Bloch theorem), while the third cumulant evaluates through a momentum integral of the same type as the occupation term of Eq.~(\ref{eq:SMmaster}) at $\omega=0$, giving Eq.~(\ref{eq:S3}); in one dimension the analogous integral gives for $S_3^{\mathrm{eq}}$ the coefficient $(3/16)\,\xi_{\mathrm{GL}}\,(u+\Xi/D)/\epsilon^{2}$ per unit length with equal contributions of $u$ and $\Xi/D$ and no contribution of $\Lambda$. Mode-coupling corrections (quartic GL terms, including the momentum-odd vertex) enter at relative order $\mathrm{Gi}/\epsilon$ and modify the far tails of the distribution.

\emph{Driven counting statistics.} The same Gaussian structure gives the generating function of the driven ensemble, from which Eqs.~(\ref{eq:Nyquist}), (\ref{eq:S3}), and (\ref{eq:CGF}) are obtained in a unified way. Along the characteristic $\bm k(t)=\bm q+2e\bm Et$ the counting weight $e^{\chi\int J|\Delta|^2dt}$ keeps the propagator of each mode Gaussian, and its inverse width $\mathcal A(\bm k)$ obeys the Riccati equation
\begin{align}
&2e\bm E\cdot\partial_{\bm k}\mathcal A=2\lambda\bar n\,\mathcal A^2+2\lambda\,\mathcal A+\chi J,
\nonumber\\
&\lim_{\mathcal T\to\infty}\frac{\ln\langle e^{\chi W}\rangle}{\mathcal T}=-\!\int_{\bm k}2\lambda\bar n\,\mathcal A,
\label{eq:riccati}
\end{align}
integrated from the downstream end of each characteristic with the equilibrium fixed point $\mathcal A^*=[\sqrt{\lambda^2-2\lambda\bar n\chi J}-\lambda]/2\lambda\bar n$ as initial condition. At $E=0$ the fixed point reproduces Eq.~(\ref{eq:CGF}) with $\chi=i\eta$; at $E\neq0$ the expansion in $\chi$ reproduces the kinetic-equation current through $O(E^2)$ and the three terms of Eq.~(\ref{eq:SMmaster}) at $\omega=0$, hence Eq.~(\ref{eq:Nyquist}) and $S_3^{\mathrm{eq}}=2T\partial_E\delta S_{\mathrm{odd}}(0)$. Direct integration of Eq.~(\ref{eq:riccati}) confirms the global detailed-balance property $\mathcal F(-E/T;E)=0$ of Ref.~\cite{Forster:2008} and the symmetry $\mathcal F(\chi;E;\bm h)=\mathcal F(-\chi-E/T;E;-\bm h)$ of Ref.~\cite{SaitoUtsumi:2008}. The second symmetry holds only if $\bm h$ is reversed, as expected for couplings odd in the field.

\section*{S5. Beyond the paraconductivity channel}

The results of the main text refer to the Aslamazov-Larkin contribution. The Maki-Thompson channel couples the fluctuation propagator to the Cooperon at the same total momentum; the common helical shift can be removed from both by a change of variable, so that only the reduced couplings $u_{\rm eff}$ and $\Xi_{\rm eff}$ enter there as well, and by analogy with the anomalous MT noise of Ref.~\cite{Kwak:2023} its nonreciprocal component is expected to be enhanced by $\tau_\phi/\tau_{\mathrm{GL}}$ and to have the characteristic frequency $\tau_\phi^{-1}$. In the slow-relaxation regime $\tau_R\gg\tau_{\mathrm{GL}}$ relevant to transition-edge devices, conductivity fluctuations adiabatically follow distribution fluctuations \cite{Kwak:2023}; the fluctuation-MCA susceptibility then acquires the same slow correlations, suggesting a correlated nonreciprocal noise component enhanced by $T_c\tau_R$. Controlled evaluation of both channels is left for future work.

\end{document}